\documentclass[prl,aps,superscriptaddress,reprint,twocolumn]{revtex4-1}
\usepackage{nn}
\usepackage{graphicx} 
\usepackage{xfrac}
\usepackage[normalem]{ulem}

\newcommand{\oparg}{\bullet}

\makeatletter
\newcommand\lettersection{\@startsection{section}{2}{\z@}%
                                     {-3.25ex\@plus -1ex \@minus -.2ex}%
                                     {-1em}%
                                     {\itshape}}
\makeatother

\renewcommand{\spin}[2]{\op{\sigma}^{#1}_{#2}}
\renewcommand{\lvl}{\mathbb{L}}
\renewcommand{\lrangle}[1]{\left\langle #1 \right\rangle}
\newcommand{\D}[1]{{#1}^{\dag}}

\newcommand{\be}{\begin{equation}}
\newcommand{\ee}{\end{equation}}
\newcommand{\bea}{\begin{eqnarray}}
\newcommand{\eea}{\end{eqnarray}}

\begin{document}

\newcommand{\documenttitle}{Beyond mean-field dynamics of the Dicke model with non-Markovian dephasing}
\title{\documenttitle}

\author{Anqi Mu}
\thanks{These authors contributed equally to this work}
\affiliation{Department of Physics, Columbia University, NY 10027, USA}
\author{Nathan Ng}
\thanks{These authors contributed equally to this work}
\affiliation{Department of Chemistry, Columbia University, NY 10027, USA}
\author{Andrew J.\ Millis}
\affiliation{Department of Physics, Columbia University, NY 10027, USA}
\affiliation{Center for Computational Quantum Physics, Flatiron Institute, New York, New York 10010, USA}
\author{David R.\ Reichman}
\affiliation{Department of Chemistry, Columbia University, NY 10027, USA}

\date{\today}

\begin{abstract}
    We present a density matrix-based time dependent projection operator  formalism to calculate the beyond mean-field dynamics of systems with non-Markovian local baths and one-to-all interactions. Such models encapsulate the physics of condensed phase systems immersed in optical cavities. We use this method, combined with tensor network influence functionals, to study the dynamics of the Dicke model coupled to non-Markovian local dephasing baths at zero temperature, which has a superradiant phase transition in the mean-field limit. The method corrects a spurious initial state dependence found in the mean-field dynamics and describes the emergence of new time scales which are absent in the mean-field dynamics. Our formalism, based on density matrices, is applicable to other quantum optical systems with one-to-all interactions at finite temperatures. 
\end{abstract}

\maketitle

\lettersection{Introduction---} The strong coupling between light and matter in an optical cavity can give rise to new quasiparticles: molecular polaritons \cite{Mandal2023,Ribeiro2018,Ebbesen2023,Xiang2024,Bhuyan2023,Xu2023}. These have attracted much attention due to the possibility of modifying chemical reactions \cite{Hutchison2012,Dunkelberger2016,Mandal2019,PannirSivajothi2022}, enabling energy transfer \cite{Coles2014,Reitz2018,Zhong2016,Xiang2020} and undergoing room temperature condensation \cite{KnaCohen2010,Plumhof2013,Daskalakis2014,Grant2016}. Modelling physically relevant molecular polariton systems requires taking into account the global coupling to the photon field as well as local baths, namely the individual vibrational environment of each molecule \cite{Herrera2017,Zeb2017}, which is typically structured and non-Markovian \cite{Chin2013,IlesSmith2016,delPinoprb2018,delPinoprl2018}. The combination of a large number of molecules together with the non-Markovian environments makes modelling such scenarios challenging.

One widely used approach is to work in the mean-field (MF) approximation, which assumes the total density matrix is a product state over the photon and the $N$ sites (representing each molecule with its local bath) \cite{FowlerWright2022,Fux2024,Keeling2025}. The mean-field approximation may be justified in the limit in which $N$ tends to infinity~\cite{Mori2013}. However, for finite $N$, we would expect correlations between sites to be important and thus lead to deviations from MF results. Existing approaches to capture $1/N$ corrections, including cumulants \cite{Kirton2017,RubiesBigorda2023,Kira2008} and path integrals \cite{DallaTorre2013,DallaTorre2016,Yulia2020}, are so far limited to the case where the local baths are modeled as Markovian. A flexible way to include finite $N$  corrections incorporating an accurate treatment of structured local baths is still lacking.

In this paper we provide a density matrix-based method to study the beyond mean-field dynamics of quantum optical systems with local baths using time dependent projection operators \cite{WillisPicard1974,WillisPicard1977,degenfeld2014self,DegenfeldSchonburgThesis}, which can naturally be combined with numerical methods like tensor network influence functionals \cite{Link2024,Nguyen2024}, to accurately treat the structured environment.
We apply this approach to the Dicke model \cite{Hepp1973,Wang1973,Dimer2007,Kirton2018,Dicke1954,Brandes2003} of a single cavity photon coupled to two-level systems, which we notate as spins, coupled to non-Markovian local dephasing baths at zero temperature. In this model, the MF dynamics displays spurious initial state dependence. Our beyond mean-field formalism resolves this pathology of initial state dependence and gives new time scales which are absent in the mean-field limit.

\lettersection{Formalism---}
We start from a one-to-all coupling type Liouville equation,
\begin{align}
    \frac{d}{dt} \op{\rho}(t) &= -i \left(\lvl_0 + \sum\limits_{i=1}^{N} \lvl_i + \lvl'_i \right) \op{\rho}(t),
\end{align}
where $\op{\rho}(t)$ is the total density matrix at time $t$. 
The Liouvillians $\lvl_0,\lvl_i,\lvl_i'$ respectively correspond to the cavity field, site $i$ and the light-matter coupling between site $i$ and the cavity field.

Under the MF approximation \cite{FowlerWright2022}, the total density matrix is fully factorized for all time, $\op{\rho} = \op{\rho}_0\otimes \bigotimes_{i=1}^N\op{\rho}_i$, where $\op{\rho}_0$ and $\op{\rho}_i$ are the density matrices for the photon and for site $i$, respectively. We then obtain the following coupled equations,
\begin{subequations}\label{eqs:mf-general}
    \begin{align}
       \frac{d}{dt} \op{\rho}_0(t) &= -i \lvl_0(t) \op{\rho}_0(t), \\
    \frac{d}{dt} \op{\rho}_{i}(t) &= -i \lvl_i(t) \op{\rho}_{i}(t), 
    \end{align}
\end{subequations}
where $\lvl_0(t)=\lvl_0+\sum_i^N\Tr_i(\lvl_i' \op{\rho}_i(t))$ and $\lvl_i(t)=\lvl_i+\Tr_0(\lvl_i' \op{\rho}_0(t))$ are the modified Liouvillians of the photon/site $i$ considering the mean-field effects of the $N$ sites/photon.

To go beyond the MF equations (\newnnref{eqs:mf-general}), we use a time-dependent generalization of the projection operator technique~\cite{WillisPicard1974, WillisPicard1977,degenfeld2014self,DegenfeldSchonburgThesis}. The projection operator $\supop{P}$ defines a splitting of the full density matrix into relevant ($\op{\rho}_{\text{rel}}=\supop{P}\op{\rho}$) and irrelevant ($\op{\rho}_{\text{irrel}} = (\supop{1}-\supop{P})\op{\rho}$) parts. An exact Nakajima-Zwanzig equation for $\op{\rho}_{\text{rel}}$ can be derived by incorporating the effects of $\op{\rho}_{\text{irrel}}$ into a memory kernel. By a prudent choice of projection operator, a low order approximation of the memory kernel can capture the essential physics.
If we project onto an uncorrelated state, $\op{\rho}_{\text{rel}}(t)=\supop{P}(t)\op{\rho}(t)=\op{\rho}_{0}(t)\otimes \bigotimes_{i=1}^N\op{\rho}_i(t)$, for models like the Dicke model at finite $N$, it can be shown that a low order expansion of the memory kernel using this projector does not give $N$-dependent unscaled steady state photon numbers~\cite{DegenfeldSchonburgThesis}, since the correlation between sites is important. For more details, see the Supplementary Material~\setsupp{See URL for the Supplementary Materials, which contains references~\cite{Konenberg2014, Argyres1964,johansson2012qutip,Lindoy2025}, detailing the critical properties of the mean-field transition;  an outline of the derivation of the beyond mean-field equations of motion;  analysis of the rise time; self-consistency of the cluster expansion; and a demonstration of the inadequacy of using a naive projection operator within the Born approximation for the beyond-mean-field dynamics.}.

To improve upon this limitation, we take a projection operator $\supop{P}(t)$ (detailed expression in~\cite{Note999}) that incorporates two-site correlations expressed via a two-site density matrix $\hat{\rho}_{ij}$. Suppressing time arguments for brevity, we have,
\begin{align}
 \begin{split}
    \op{\rho}_{\text{rel}} &\equiv \op{\rho}_{0} \otimes \Big[ \bigotimes\limits_{i=1}^{N} \op{\rho}_i + \sum\limits_{\substack{j < i}}^N (\op{\rho}_{ij} - \op{\rho}_{i}\otimes\op{\rho}_{j}) \otimes \bigotimes\limits_{k\neq i, j}^N \op{\rho}_k \Big] \\
     &\equiv \op{\rho}_0 \otimes \op{\rho}_{N,\text{rel}}.
 \end{split}
\end{align}
In this expression the density matrix is expressed as a product of a photon part $\hat{\rho}_0$ and a site-only part $\op{\rho}_{N,\text{rel}}$ given by the first two terms of a cluster expansion. Correlations between the cavity and the sites are accounted for within a time-nonlocal memory kernel. 

We make an expansion of the memory kernel to second order in the light-matter coupling (as we will see this gives corrections of order $1/N$) and obtain the following coupled equations,
\begin{widetext}
\begin{subequations}\label{eqs:bmf-general}
    \begin{align}
        &\frac{d}{dt} \op{\rho}_0(t) = -i \lvl_0(t) \op{\rho}_0(t) -\int_0^{t} \!\!dt' \, \Tr_{\overline{0}} \left[ \sum\limits_{i=1}^{N} \Delta\lvl'_i(t) \supop{G}(t,t') \supop{Q}(t') \sum\limits_{j=1}^N \Delta\lvl'_j(t') \op{\rho}_{\text{rel}}(t')\right], \\
    &\frac{d}{dt} \op{\rho}_{ij}(t) = -i (\lvl_i(t)+\lvl_j(t)) \op{\rho}_{ij}(t)- \int_0^t \!\!dt' \, \Tr_{\overline{i,j}} \Tr_0 \left[ (\Delta\lvl'_i(t) + \Delta\lvl'_j(t)) \supop{G}(t,t') \supop{Q}(t') \sum\limits_{k=1}^N \Delta\lvl'_k(t')  \op{\rho}_{\text{rel}}(t')\right],\\
    &\supop{G}(t,t')=\tord \exp \left[ -i \int_{t'}^t \!\!dt'' \, (\lvl_0(t'')+\sum_{i=1}^N\lvl_i(t''))\right],
    \end{align}
\end{subequations}
\end{widetext}
where $\Delta \lvl'_i(t)=\lvl'_i-\Tr_i(\lvl_i' \op{\rho}_i(t))-\Tr_0(\lvl_i' \op{\rho}_0(t))$ captures fluctuations around the MF dynamics and $\supop{Q}(t)=1-\supop{P}(t)$. The subscript $\overline{0}$ refers to only the $N$ sites and $\overline{i,j}$ means the complement of sites $i$ and $j$ among the $N$ sites. Though \newnnref{eqs:bmf-general} involves $\op{\rho}_0$, which has a large Hilbert space, as we will show in our example in the next section, our equations only require at most the second moments of $\op{\rho}_0$ (while MF only requires, at most, the first moments). Finally, the effect of non-Markovian local baths on the evolution of $\op{\rho}_{ij}$ can be easily incorporated through methods for open quantum systems that make use of an efficient bath representation, namely pseudomode~\cite{Garraway1997, Dorda2014, Tamascelli2018, Lambert2019, Pleasance2020, Park2024, Debecker2024} and influence functional-based~\cite{Tanimura1989, Makri1995a, Shi2018, Strathearn2018, Link2024} approaches.

The mathematical form of this projector $\supop{P}(t)$ takes
\begin{align}
    \begin{split}
        \supop{P}(t) \oparg &= \left( \Tr_{\overline{0}} \oparg \right) \otimes \op{\rho}_{N,\text{rel}}(t) \\
        &\phantom{=} + \op{\rho}_0(t) \otimes \left( \supop{P}_N(t) \Tr_0 \oparg \right) \\
        &\phantom{=} - \left( \op{\rho}_0(t) \otimes \op{\rho}_{N,\text{rel}}(t) \right) (\Tr \oparg),
    \end{split} 
\end{align}
where $\supop{P}_N(t)$ projects the density matrix for all sites $\Tr_0 \op{\rho}(t)$ onto $\op{\rho}_{N,\text{rel}}(t)$ and the expression of $\supop{P}_N(t)$ is given in the Supplement~\cite{Note999}.

\lettersection{Model---} We apply the formalism of \newnnref{eqs:bmf-general} to the Dicke model of $N$ spins-$\tfrac{1}{2}$ with local dephasing baths, 
We also add a Markovian photon loss term with rate $\kappa$. The Hamiltonian for this model is given by $\op{H} = \op{H}_0 + \sum_i \op{H}_i + \op{H}'_i$ where,
\begin{subequations}
    \begin{align}
        \op{H}_0 = \Omega &\D{\op{a}}\op{a}, \qquad\qquad
        \op{H}_i = \omega_z \spin{z}{i}+\op{H}_i^E, \\
        &\op{H}'_i = (g/\sqrt{N}) (\op{a} + \D{\op{a}})\spin{x}{i},
    \end{align}
\end{subequations}
and so correspond to the Liouvillians,
\begin{subequations}
    \begin{align}
        \lvl_0 \op{\rho} &= [\op{H}_0, \op{\rho}] + i \kappa \left( \op{a}\op{\rho}\D{\op{a}} - \acomm{\D{\op{a}}\op{a}}{\op{\rho}}/2 \right), \\
        \lvl_i &\op{\rho} = [\op{H}_i, \op{\rho}], \qquad\qquad 
        \lvl'_i \op{\rho} = [\op{H}'_i, \op{\rho}].
    \end{align}
\end{subequations}
Here $\Omega$ and $\omega_z$ are respectively the cavity and spin frequencies, and $g$ is the light-matter coupling strength. In this work, all energy scales will be given in units of $\Omega$. As in preceding works \cite{DallaTorre2016,Kirton2017}, we ignore the diamagnetic $A^2$ term.
Our results would be appropriate for effective models without the $A^2$ term, which can be realized in various experimental platforms \cite{Kono2025,Baumann2010Dicke}. The results would differ in gauge-invariant QED systems, though there are works showing that it's justified to ignore the $A^2$ term \cite{Vukics2012,Domokos2014} in related settings.
Each spin couples to a harmonic dephasing bath through
\begin{align}
\op{H}_{i}^E=\sum_j\Big[\nu_j\op{b}_j^{\dagger}\op{b}_j+\xi_j(\op{b}_j+\op{b}_j^{\dagger})\spin{z}{i}\Big],
\end{align}
where $\xi_j$ is the coupling strength between $j$th bosonic mode and the spin. We assume the bath parameters and couplings are the same for each spin. The associated bath spectral density is $J(\omega) = \sum_j \xi_j^2 \delta(\omega - \nu_j)$,
which is taken to have the continuous form $J(\omega) = (\alpha / 2) \omega_c (\omega/\omega_c)^s \exp(-\omega/\omega_c)$. The cutoff frequency $\omega_c$ sets the timescale for the bath, while $\alpha$ controls the strength of the bath coupling. For concreteness, we will focus on Ohmic baths and take $s = 1$, though our methods hold also for $s \neq 1$. Throughout this work, we take these dephasing baths to be initially held at zero temperature, factorized from the spin state. Additionally, we will fix $\omega_c = 1$ and $\alpha = 0.3$, along with $\omega_z = 0.025$ and $\kappa=1$. This model has a $\supop{Z}_2$ symmetry under $a\rightarrow -a,\sigma_i^x\rightarrow-\sigma_i^x$.  In the thermodynamic limit, absent the local baths, there is a superradiant (SR) phase transition from a normal phase with $\lrangle{\op{a}}/\sqrt{N}=0$ to a symmetry broken phase with $\lrangle{\op{a}}/\sqrt{N}\neq 0$ at critical coupling $g_c^2=\omega_z(\Omega^2+\kappa^2)/(2\Omega)$ \cite{Kirton2018}. We first look at the mean-field dynamics.

\lettersection{Mean-field limit---}
From \newnnref{eqs:mf-general} and assuming permutation symmetry of all sites, we see that the mean-field dynamics are exactly captured by an effective model describing the evolution of a single spin-boson system subjected to a time-dependent external field $a(t) \equiv \langle \op{a}(t) \rangle/\sqrt{N}$ and a corresponding evolution equation for the photon field $a(t)$, 
\begin{subequations}\label{eqs:mf-eom}
\begin{align}
    \tfrac{d}{dt} a(t) &= (-i \Omega - \kappa) a(t) - i g \Tr_i (\spin{x}{i} \op{\rho}_i ), \\
    i \tfrac{d}{dt} \op{\rho}_i(t) &= \comm{\omega_z\spin{z}{i}+ (2 g \Re a(t)) \spin{x}{i} + \op{H}_i^E }{\op{\rho_i}(t)}\\
    &\equiv [\op{H}_{\text{eff}}(t),\op{\rho}_i(t)]. \nonumber
\end{align} 
\end{subequations}
We can account for non-Markovianity in numerical simulations through the time-evolving matrix product operator (TEMPO) method~\cite{Strathearn2018, FowlerWright2022, FowlerWrightThesis}.
In particular, we make use of its time-translationally invariant~\cite{Link2024} variant, which reduces the computational cost required to propagate the dynamics to long times. We use $0.04$ for the timestep and $10^{-8.5}$ for the relative threshold of SVD truncation for TEMPO.

\begin{figure}
    \centering
    \includegraphics[width=\columnwidth]{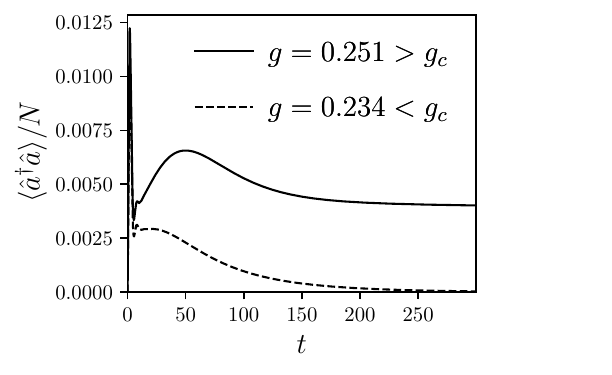}
    \caption{
    Time evolution of the mean-field scaled photon occupation in the open Dicke model with zero temperature local Ohmic dephasing baths starting from an initial state in which the cavity is initially in the vacuum state while the spins are initially polarized in the $\hat{x}-\hat{z}$ plane at an angle of $\theta = 0.825\pi$ from the $+\hat{z}$ direction, for  two values of light-matter coupling, one (solid) such that the system is superradiant (solid)  and the other normal (dashed).  The transition is estimated to occur at $g_c \approx 0.246$.
    }
    \label{fig:mf-tempo}
\end{figure}

It was shown by Kirton and Keeling that SR is fully suppressed~\cite{Kirton2017} when the local dephasing is taken to be Markovian.  However we find (see \newnnref{fig:mf-tempo})  that, consistent with Fowler-Wright~\cite{FowlerWrightMasters}, for non-Markovian dephasing a SR transition occurs as the coupling is increased in MF theory. The uniqueness of pure Markovian dephasing can be seen in its Linbladian, $\Gamma_{\phi} (\spin{z}{} \op{\rho} \spin{z}{} - \op{\rho})$, which contains neither a response (imaginary part) nor a reciprocal term encoding a detailed balance relation. 
As shown in the Appendix, this can only be consistent with infinite temperature of the bath.
This underscores the importance of carefully modeling the temporal behavior of the bath.

We provide an analytical estimation of the steady state phase boundary $g_c$ based on linear response theory.
By treating the light-matter coupling as a perturbation on top of the pure dephasing model in $\op{H}_{\text{eff}}$ (\newnnref{eqs:mf-eom}), and combining with the steady state condition~\cite{Note999,FowlerWrightMasters}, we obtain (cf.\ \cite{DallaTorre2016, Kirton2017})
\bea
g_c^2&=&\left(\Omega^2+\kappa^2\right)/(-2\Omega\chi),
\label{Eq:linear response}
\eea
where $\chi=-\int_0^{\beta_T} d\tau \langle \sigma^x(\tau)\sigma^x\rangle$ is the spin-spin correlation function of the pure dephasing model evaluated at temperature $1/\beta_T$~\cite{WeissBook}.

\begin{figure}
    \centering
    \includegraphics[width=\columnwidth]{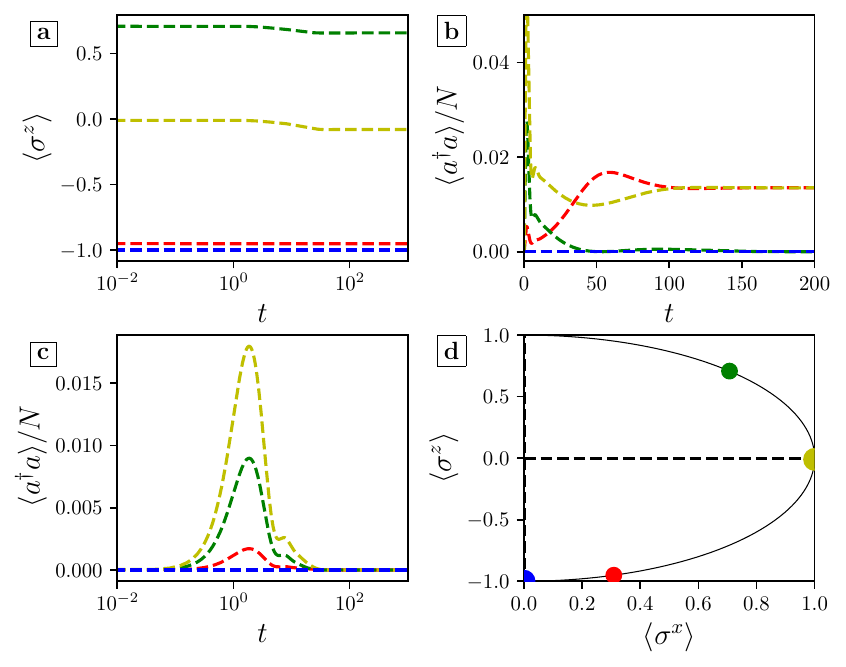}
    \caption{
    Mean-field dynamics of the open Dicke model with local zero temperature Ohmic dephasing baths. The cavity is initially in the vacuum state. The four initial spin states are labeled in panel (d) (position on the Bloch sphere within the x-z plane). Under MF, spin initially pointing along $-z$ is a trivial fixed point when the initial photon occupation is zero. \textbf{(a)} and \textbf{(c)} Dynamics of the imbalance $\langle \spin{z}{} \rangle$ and photon number for $g = 0.16 < g_c$. We can clearly see that different initial states end up with different values of $\langle \spin{z}{} \rangle$, with their dynamics becoming frozen around the time when the photon number drops to negligibly small values. \textbf{(b)} Dynamics of the scaled photon number in the putative SR phase, for $g = 0.28$ slightly larger than $g_c$. However, the initial state dependence still persists.
    }
    \label{fig:mf-state-dependence}
\end{figure}

\nnref{fig:mf-state-dependence}{Figure~}{} reveals an initial state dependence of the long time limit  of the mean-field dynamics. In the normal phase, we can see that even though $n(t) \equiv |a(t)|^2$ vanishes at long times, different initial conditions lead to different values of $\langle \spin{z}{} \rangle$. This initial state dependence is most striking in the region slightly above $g\gtrsim g_c$, for which some initial conditions show the absence of superradiance ($n(t) \to 0$) at long times while for others the system evolves to the superradiant state. As the coupling $g$ further increases all initial conditions eventually reach a unique superradiant state. This can be can be understood as a pathology of mean-field theory. When $a(t\to\infty) = 0$, in \newnnref{eqs:mf-eom} there is a conserved quantity, $\spin{z}{}$. The MF Dicke model under local dephasing may not fully thermalize; that is, starting from different initial conditions, the steady state density matrix $\op{\rho}_{\text{eff}}$ can be different as long as it satisfies $[\op{\rho}_{\text{eff}},\spin{z}{}]=0$. Thus, in the normal phase, different steady state values of $\langle \spin{z}{} \rangle$ may be possible.
This happens because in MF we replace $\op{a}$ by $\langle \op{a}\rangle$ and ignore the order 1 fluctuations in $\langle \op{a} \rangle$. Away from the MF limit, that is, in finite $N$ systems, $\hat{\sigma}^z$ is prevented from becoming exactly conserved by the nonvanishing light-matter coupling term $H_i'$, which is of order $O(g/\sqrt{N})$. Since one spin can interact with another spin via the photon, which is a process with a matrix element of order $O((g/\sqrt{N})^2)$, this leaves open a mechanism by which a spin can thermalize with its surrounding spins on a timescale of $O((\sqrt{N}/g)^2)$, a timescale which formally does not exist in the MF limit.

\lettersection{Beyond mean-field---} We now show how our beyond mean-field dynamics approach leads to a unique steady state and reveals the $O((\sqrt{N}/g)^2)$ timescale. Using the aforementioned formalism (\newnnref{eqs:bmf-general}) we expand the memory kernel to second order in the light-matter coupling $O(g^2/N)$ and obtain a set of coupled integro-differential equations for $\op{\rho}_{ij}$ and first- and second-moments of $\op{\rho}_{0}$, where it can be explicitly seen that $\spin{z}{}$ is no longer conserved due to the presence of two-site correlations, as expected~\cite{Note999}. The time non-locality within these equations is dampened by the presence of photon loss, which attenuates the contributions as $\exp(-\kappa (t-t'))$.

\begin{figure}
    \centering
    \includegraphics[width=\columnwidth]{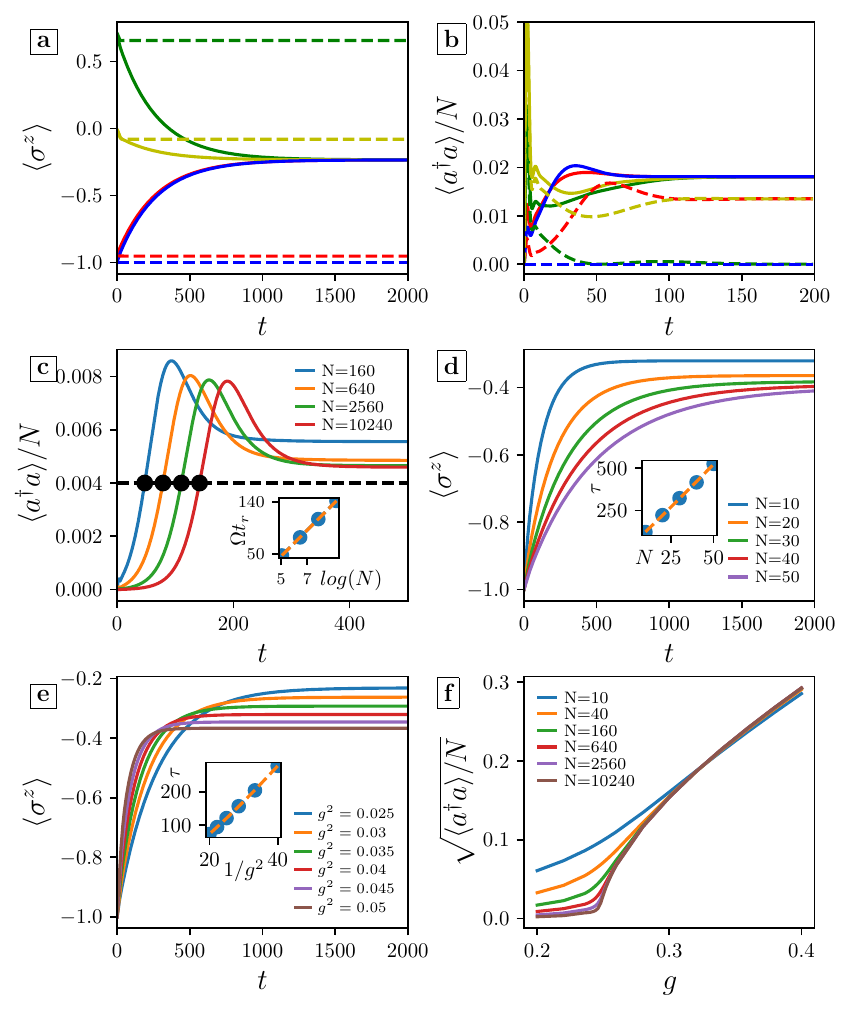}
    \caption{\textbf{(a)} Dynamics of the imbalance $\langle \spin{z}{} \rangle$ for $g = 0.16 < g_c$ and $N=10$ with the same four initial conditions and the same set of parameters as in \newnnref{fig:mf-state-dependence}. Dashed lines correspond to mean-field results. \textbf{(b)} Dynamics of the scaled photon number for $N=10$ and $g = 0.28 > g_c$ under the same conditions as (a). Dashed lines correspond to mean-field results. \textbf{(c)} Dynamics of the scaled photon number for $g = 0.26 > g_c$ and different $N$. The dashed line roughly corresponds to the half maximum of these curves, and the intersections (black circles) give a measure of the rise time $t_r$, which the inset shows increases as $\log N$. \textbf{(d)} Relaxation of the imbalance at long times for fixed $g$ and varying $N$ for $g = 0.2 < g_c$, showing that the relaxation time $\tau$ grows linearly with $N$ (inset). \textbf{(e)} Same as in (d), but for fixed $N = 10$ and varying $g$, showing that $\tau \propto 1/g^2$ (inset). \textbf{(f)} Steady state photon number as a function of $N$ obtained from the same initial condition as in (c,d,e), where the cavity is unoccupied and the spins are fully polarized in the $-\hat{z}$ direction. Panels (d) and (e) shows that the relaxation time scales as $N/g^2$ in the normal phase.
    }
    \label{fig:bmf}
\end{figure}

We first apply this approach to resolve the issue of initial state dependence in mean-field theory. As shown in \newnnref[(a,b)]{fig:bmf}, using the same set of parameters and initial states as in \newnnref{fig:mf-state-dependence}, we find that while the dynamics at early times follows that of the mean-field (shown in dashed lines), at long times all of the different curves converge to a common value, which suggests that the desired thermalization mechanism has been restored. 

Let us comment on these two temporal regimes, starting with the behavior at early times.
We show in \newnnref[c]{fig:bmf} the evolution of the scaled photon number from an initial condition in which the cavity mode is unoccupied and all the spins are fully polarized in the $-\hat{z}$ direction, for $g > g_c$.
With pure dephasing baths, this initial condition has trivial dynamics in the MF limit, in that the spins and the cavity remain stuck in their initial states for all times.
Away from the MF limit at early times, the two-site correlations start to establish and the scaled photon number $n \equiv \langle \D{\op{a}}\op{a} \rangle/N$ increases from $n=0$ to $n\sim O(1/N)$ around $t\approx 1$ (see~\cite{Note999}). 
After this, the mean-field ($N$-independent) processes take over and lead to an exponential growth of $n$, so that we have $n \sim O(1/N) e^{\gamma t}$ for some $N$-independent rate $\gamma$. This process ends when $n \sim O(1)$, from which we obtain the observed initial timescale that increases as $\log N$. 

At the other extreme, at late times for $g < g_c$, we find that the convergence towards the beyond mean-field steady state occurs on a timescale $O(N/g^2)$.
In this long-time relaxation regime the state of the spins is already far from being factorized.
Thus the mechanism underlying the $O(N/g^2)$ timescale is slightly different, and most clearly emerges from the time-nonlocal equations in our projection operator approach.
At late times when the state of the system is close to the steady state, the time-nonlocal memory kernels become approximately time-translationally invariant.
Moreover, as these memory kernels decay very quickly on timescales of $O(1/\kappa)$ (see~\cite{Note999}), the evolution of the state becomes approximately Markovian with a rate set by the magnitude of the memory kernel, which is proportional to $g^2/N$.
This then leads to the slow relaxation we observe, which we show for $\langle \spin{z}{} \rangle$ in \newnnref[(d,e)]{fig:bmf}.
In panels (d) and (e) of this figure, we show that the observed relaxation timescale is proportional to both $N$ and $1/g^2$, respectively.

With the knowledge of the relevant timescales in the beyond mean-field dynamics, we show the steady state value of $\sqrt{n}$ across different system sizes (\newnnref[f]{fig:bmf}).
In the MF limit this quantity is exactly the order parameter $\langle \op{a} \rangle/\sqrt{N}$.
We see that the sharpening of the crossover with increasing $N$ is captured within our projection operator approach.

Finally, we note that our approach does not provide a complete treatment of finite $N$ systems. We find symmetry-broken steady states for sufficiently large $N$ when starting from initial state that does not respect the $Z_2$ symmetry, whereas on general grounds we expect that a finite-$N$ system should not break a symmetry. 
This is reminiscent of results from cumulant expansions for central spin models with Markovian local baths~\cite{fowler2023determining} and results from the inability (common to both cumulant expansion and our projection operator approach) to capture the instanton processes (formally $\mathcal{O}(e^{-N}$) that restore the $Z_2$ symmetry). Inclusion of these processess is an important goal for future work.

\lettersection{Discussion---}
In this Letter we present a formalism to include finite $N$ corrections to the dynamics of systems with one-to-all type interactions. 
We apply it to the Dicke model in the presence of zero temperature local dephasing baths, where non-Markovian effects are unavoidable, and compare the results against MF dynamics.
We find initial state dependence in the MF dynamics, which we attribute to the neglect of fluctuations assumed by mean-field theory, and show that it is amended by the beyond mean-field dynamics. We also obtain information about the phase transition as well as characterize different timescales governing the dynamics. We believe this method also applies to other quantum optical systems with one-to-all interactions, even with finite temperature local baths. We also note a recently proposed wavefunction based method (CUT-E) \cite{PerezSanchez2023,PerezSanchez2025} that also captures non-Marovian effects with finite $N$ corrections. However so far the CUT-E method has been formulated for the zero temperature case and first excitation manifold. We will give a more thorough analysis of such beyond mean-field methods in a later work.

\lettersection{Acknowledgments---}
We thank Jonathan Keeling for insightful discussions. 
This work was performed with support from the U.S.\ Department of Energy, Office of Science, Office of Advanced Scientific Computing Research, Scientific Discovery through Advanced Computing (SciDAC) program, under Award No. DE-SC0022088.
This research used resources of the National Energy Research Scientific Computing Center (NERSC), a U.S. Department of Energy Office of Science User Facility located at Lawrence Berkeley National Laboratory, operated under Contract No. DE-AC02-05CH11231.
The Flatiron Institute is a division of the Simons Foundation.

\setcounter{equation}{0}
\renewcommand{\theequation}{A\arabic{equation}}

\bibliography{references}

\section*{End Matter}

\begin{center}
    \it Appendix A: Markovian dephasing as a high temperature limit
\end{center}
The Lindblad evolution of a qubit with a single jump operator $\spin{z}{}$ leads to the time-discretized influence functional~\cite{Makri1995a},
\begin{align}
    \begin{split}
        I[\{s^{\pm}_n\}] &= \exp \Big[ - \sum_{i\geq j} (s^+_i - s^-_i) (\eta_{i-j} s^+_j - \eta_{i-j}^* s^-_j )\Big],
    \end{split} \nonumber \\
    \Re \eta_{k} &= \delta_{0, k} \Gamma_{\phi} \Delta t / 2.
    \label{eq:app-eta}
\end{align}
From this, we seek to recover the bath spectral density $J(\omega)$ that generates this dynamics in the same vein as in Appendix C.F of \cite{Ivander2024}.
Note that this procedure assumes that the system is coupled to a harmonic bath initially prepared at some temperature $1/\beta_T$ independently of the system.

Since $\Re \eta_{k\neq 0} = 0$, we must have,
\begin{align}
    \Re \eta_k &=  \!\! \int\limits_{-\infty}^{\infty} \!\!d\omega \, \frac{(\text{sgn} \omega)J(|\omega|)}{\tanh(\beta_T\omega/2)} \left[\frac{\sin\tfrac{\omega \Delta t}{2}}{\omega} \right]^2 e^{-i k \Delta t \omega}.
\end{align}
In the continuous time limit, we recover a spectral density independent of $\Delta t$,
\begin{align}
    J(\omega) &= \frac{\Gamma_{\phi}}{\pi} \tanh{(\beta_T\omega/2)}.
\end{align}
This spectral density leads to a bath with an infinite reorganization energy $\Lambda \propto \int_0^{\infty} \frac{d\omega}{\omega} J(\omega)$ due to an ultraviolet divergence.
To make this model physical, we introduce an upper frequency cutoff $\omega_c$, $\tilde{J}(\omega) = J(\omega) u(\omega/\omega_c)$, where the cutoff function $u(x)$ is such that $u(0) = 1$ and $u(x\to\infty) = 0$.
This gives $\Lambda \propto \Gamma_{\phi} f(\beta_T \omega_c)$, with $f(z) = \int_0^{\infty} \frac{dx}{x} \tanh(z x/2) u(x)$.
By \newnnref{eq:app-eta}, we require $\omega_c\to\infty$ to enforce $\Re \eta_{k\neq 0} = 0$.
The only way that this limit can be taken so as to maintain a finite reorganization energy is to have $\beta_T$ decrease no slower than $1/\omega_c$.
Thus we see that only at infinite temperature would we obtain a physically meaningful harmonic bath that produces Lindblad dynamics, i.e.~\newnnref{eq:app-eta}.

\newpage

\pagestyle{plain}
\makeatletter
\renewcommand{\c@secnumdepth}{0}

\setcounter{page}{1}
\setcounter{equation}{0}
\setcounter{section}{0}
\setcounter{figure}{0}
\renewcommand{\theequation}{S\arabic{equation}}
\renewcommand{\thesection}{SM-\Roman{section}}
\renewcommand{\thefigure}{S\arabic{figure}}
\onecolumngrid

\begin{center}
\large\bf Supplementary Material for ``\documenttitle''
\end{center}

\vspace*{2em}

This supplement is divided into \ref*{SM:num-parts} parts, detailing:
\begin{enumerate}
    \item the critical properties of the mean-field SR transition (\newnnref{sec:SM-critical-exp}),
    \item a sketch of the derivation of the beyond mean-field equations of motion using time-dependent projection operators (\newnnref{sec:SM-bmf}),
    \item details on initial exponential growth of photon number in finite $N$ dynamics (\newnnref{sec:SM-logN-rise-time}),
    \item self-consistency of the assumption underlying the cluster expansion (\newnnref{sec:SM-hierarchical-ordering}),
    \item an explicit demonstration of the failure of using a na\"ive projection operator within the Born approximation to describe macroscopic photon occupations at the steady state limit of finite $N$ systems (\newnnref{sec:SM-naive-projector-bad}).
    \item a benchmark on the Dicke model without local baths (\newnnref{sec:SM-benchmark})
    \item a benchmark on the Dicke model with local baths modeled by a single bosonic mode with Lindbladian dissipation (\newnnref{sec:SM-benchmark-singleboson})
    \item a benchmark on the Dicke model with Ohmic local baths (\newnnref{sec:SM-benchmark-ohmicbath})
    \item and effects of finite temperature  local baths (\newnnref{sec:SM-benchmark-finiteT})
    \label{SM:num-parts}
\end{enumerate}

\section{\label{sec:SM-critical-exp}Critical properties of the mean-field transition}
As stated in the main text, the steady state of the mean-field Dicke model can be obtained by ensuring self-consistency between the steady state value of $\langle \spin{x}{} \rangle$ in the biased spin-boson model and the steady state relation between the cavity variable $a_{\text{ss}} \equiv \langle \op{a}(t\to\infty) \rangle/\sqrt{N}$ and $\langle \spin{x}{} \rangle_{\text{ss}}$.
Doing this reveals that quantities like $\Re a_{\text{ss}}$ just above the transition must scale as $\Re a_{\text{ss}} \sim (g - g_c)^{\beta}$ where $\beta = \tfrac{1}{2}$. 

To see this, note that from the steady state of \newnnref{eqs:mf-eom}, we obtain $\Re a_{\text{ss}} = (\Omega/\kappa) \Im a_{\text{ss}}$ and $\langle \spin{x}{} \rangle_{\text{ss}} = -\tfrac{1}{2\Omega}(\Omega^2 + \kappa^2)(2 g \Re a_{\text{ss}})/g^2$. Then we equate  $\langle \spin{x}{} \rangle_{\text{ss}}$ to the expansion of the steady state $\langle \spin{x}{} \rangle_{\text{ss}}^{\text{SB}}$ in the biased spin-boson model in terms of the tunnelling strength $\Delta \equiv 2 g \Re a_{\text{ss}}$, i.e., $\langle \spin{x}{} \rangle_{\text{ss}}^{\text{SB}} \approx a_0 (2 g \Re a_{\text{ss}}) + a_1 (2 g \Re a_{\text{ss}})^3$. From this, we must have that $a_0 = -(g_c^{-2})(\Omega^2 + \kappa^2)/(2\Omega)$. Therefore, when $g$ is just above $g_c$, $\Re a_{\text{ss}} \sim \sqrt{g - g_c}$. By the proportionalities stated above, $\Im a_{\text{ss}}$ and $\langle \spin{x}{} \rangle_{\text{ss}}$ must have the same critical behavior.

\subsection{Critical coupling}
We shall work with the following ideas in mind:
\begin{enumerate}
\item Analytically approaching the phase boundary from the normal phase is ill-defined, since there are infinitely many states of the TLS+bath that are stationary with respect to the steady state Hamiltonian, which is of pure-dephasing form by definition. We will therefore situate ourselves infinitesimally above $g_c$, in the superradiant phase (assuming that it exists).
\item The steady state Hamiltonian for parameters infinitesimally beyond the superradiant phase boundary describes a biased spin-boson model with a small tunnelling matrix element, $\Delta \ll 1$. We assume~\cite{Konenberg2014} that there are unique values for long time averages, which are equal to thermodynamic averages with respect to the initial temperature of the harmonic oscillator bath.
\item We assume that the polarization of the TLS along the x-direction is adequately described by perturbation theory. This is only needed for us to make analytical calculations, but is not necessary in general.
\end{enumerate}

The above assumptions allow us to work directly at the steady state, ignoring all transient dynamics.
The steady state condition on the cavity field yields the relation,
\begin{align}
\Re a_{\text{ss}} &= \frac{-g \Omega}{\Omega^2 + \kappa^2} \lrangle{\spin{x}{}}_{\text{ss}}.
\end{align}
We have also that,
\begin{align}
  \lrangle{\spin{x}{}}_{\text{ss}} &\overset{!}{=} \lrangle{\spin{x}{}(\Delta)}_{\beta} \equiv \Tr \left( \spin{x}{} \frac{\exp(-\beta H(\Delta))}{Z(\Delta)} \right), \\
  H(\Delta) &= \Delta \spin{x}{} + \underbrace{ \omega_z \spin{z}{} + \spin{z}{} \sum_j \xi_j \left( \op{b}_j + \op{b}_j^\dagger \right) + \sum_j \nu_j \op{b}_j^\dagger b_j }_{\equiv \op{H}_0},
\end{align}
where $\Delta = 2 g (\Re a_{\text{ss}})$.
These equations must be self-consistently satisfied, as we have stated in our consideration of the critical exponent.

To obtain the phase boundary, we shall make use of our previous identification of $a_0 \propto g_c^{-2}$ and calculate the thermal average perturbatively via the Dyson expansion,
\begin{align}
\lrangle{\spin{x}{}(\Delta)}_{\beta} &\approx  \Delta \left( \underbrace{-\int\limits_0^{\beta} d\tau \, \Tr \Bigg[ \frac{e^{-\beta \op{H}_0}}{Z_0} e^{\tau \op{H}_0} \spin{x}{} e^{-\tau \op{H}_0} \spin{x}{} \Bigg]}_{\equiv \chi} \right), \\
\chi &= -\int\limits_0^{\beta} d\tau \, \frac{\cosh(\beta \omega_z - 2 \tau \omega_z)}{\cosh(\beta \omega_z)} \exp \left[ - 4 \int\limits_0^{\infty} d\omega \, \frac{J(\omega)}{\omega^2} \Bigg( \coth\left( \frac{\beta \omega}{2} \right) \left( 1 - \cosh(\tau \omega) \right) + \sinh(\tau \omega) \Bigg) \right]. \label{eq:SM-spin-spin-susceptibility}
\end{align}
For the special case of zero temperature,
\begin{align}
\label{eq:SM-spin-spin-susceptibility-zero-temp}
\chi &= -2 \int\limits_0^{\infty} d\tau \, e^{-2\tau\omega_z} \exp \left[ - 4 \int\limits_0^{\infty} d\omega \, \frac{J(\omega)}{\omega^2} \left( 1 - e^{-\tau \omega} \right) \right].
\end{align}
From the above, we have that,
\begin{align}
    \chi \equiv a_0 &= -(g_c^{-2}) \frac{\Omega^2 + \kappa^2}{2\Omega}, \\
    \Longrightarrow g_c^2 &= \frac{-1}{\chi} \frac{\Omega^2 + \kappa^2}{2\Omega}.
\end{align}

\section{\label{sec:SM-bmf}Beyond mean-field dynamics using time-dependent projection operators}

We wish to define a projector such that its action on the full (mixed) state isolates the uncorrelated state of all the sites plus all two-site irreducible correlations.
However, we shall restrict the projected state to be uncorrelated between the cavity and all $N$ sites due to numerical constraints. As for our Dicke model example, since we expect $O(N)$ photon excitations in the superradiant phase, the density matrix of the cavity mode plus a single site (spin and a representation of its local bath) could become too large to represent accurately.
Hence we will rely on the memory kernel being able to account for irreducible correlations between the cavity mode and a single site.
In all, the action of the projector should give,
\begin{align}
    \begin{split}
        \op{\rho}_{\text{rel}}(t)=\supop{P}(t) \op{\rho}(t) &= \op{\rho}_0(t) \otimes \Bigg[ \bigotimes\limits_{i=1}^N \op{\rho}_i(t) + \Bigg( \sum\limits_{\substack{i,j=1 \\ i > j}}^N \left( \op{\rho}_{i,j}(t) - \op{\rho}_i(t) \otimes \op{\rho}_j(t) \right) \otimes \bigotimes\limits_{\substack{k=1 \\ k\neq i,j}}^N \op{\rho}_k(t) \Bigg) \Bigg] \\
        &\equiv \op{\rho}_0(t) \otimes \op{\rho}_{N,\text{rel}}(t).
    \end{split}
\end{align}
Adapting the construction by Willis and Picard~\cite{WillisPicard1977}, one finds that the form of this projector should take
\begin{align}
    \begin{split}
        \supop{P}(t) \oparg &= \left( \Tr_{\overline{0}} \oparg \right) \otimes \op{\rho}_{N,\text{rel}}(t) \\
        &\phantom{=} + \op{\rho}_0(t) \otimes \left( \supop{P}_N(t) \Tr_0 \oparg \right) \\
        &\phantom{=} - \left( \op{\rho}_0(t) \otimes \op{\rho}_{N,\text{rel}}(t) \right) (\Tr \oparg),
    \end{split} \\
    \begin{split}
        \supop{P}_N(t) \oparg &= \sum\limits_{\substack{i,j = 1\\i > j}}^N \left(\Tr_{\overline{i,j}} \oparg \right) \otimes \bigotimes\limits_{\substack{k=1 \\ k\neq i, j}}^N \op{\rho}_k(t) + \op{\rho}_{ij}(t) \otimes \sum\limits_{\substack{k=1 \\ k\neq i, j}}^N \left( \Tr_{\overline{k}} \oparg \right) \otimes \bigotimes\limits_{\substack{l=1 \\ l \neq i, j, k}}^N \op{\rho}_l(t) \\
        &\phantom{=} - (N-2) \sum\limits_{\substack{i,j = 1\\i > j}}^N \left(\Tr \oparg \right) \op{\rho}_{ij}(t) \otimes \bigotimes\limits_{\substack{k=1\\k \neq i,j}}^N \op{\rho}_k(t) \\
        &\phantom{=} - \left( \binom{N}{2} - 1 \right) \supop{P}_{N,\text{mf}}(t) \oparg,
    \end{split} \\ 
    \begin{split}
        \supop{P}_{N,\text{mf}}(t) \oparg &= -(N-1) \left( \bigotimes\limits_{i=1}^N \op{\rho}_i(t) \right) \left(\Tr \oparg \right) + \sum\limits_{i=1}^N \left( \Tr_{\overline{i}} \oparg \right) \otimes \bigotimes\limits_{\substack{j=1\\ j \neq i}}^N \op{\rho}_j(t).
    \end{split}
\end{align}
Here we use the notation $\overline{i,j}$ to denote the complement of sites $i$ and $j$ among the $N$ sites, i.e., $\overline{i,j} \equiv \{1, \ldots, N\} \setminus \{i,j\}$. Likewise $\overline{0}$ means all the $N$ sites and $\overline{i}$ represents all sites except site $i$.
For convenience, we have defined useful intermediate projection operators $\supop{P}_N(t)$ and $\supop{P}_{N,\text{mf}}(t)$, both of which act only on density matrices for all sites $\Tr_0 \op{\rho}$.
The former projects onto the state with at most two-site irreducible correlations, while the latter projects the same density matrix onto the completely factorized state (i.e., mean-field).
Note that all of these definitions involve time-dependent states such as $\op{\rho}_i(t)$, so that these projectors are dependent on the given {\it state}, unlike usual choices such as the Argyres-Kelley projector~\cite{Argyres1964}.
For this reason, these are termed {\it self-consistent projection operators}.

It can be checked that these operators satisfy the following important properties:
\begin{enumerate}
    \item $\Tr_{\overline{i,j}} \supop{P}_N(t) \oparg = \Tr_{\overline{i,j}} \oparg$, 
    \item $\comm{\supop{P}_N(t)}{\frac{d}{dt}} \left( \Tr_0 \op{\rho}(t) \right) = 0$, 
    \item $\Tr_{\overline{0}}\supop{P}(t)\oparg = \Tr_{\overline{0}} \oparg$, 
    \item $\supop{P}(t) \supop{P}(t') = \supop{P}(t)$ for all $t, t'$, which implies that, for $\supop{Q}(t) \equiv \supop{1} - \supop{P}(t)$, $\supop{Q}(t) \supop{Q}(t') = \supop{Q}(t')$ and $\supop{P}(t) \supop{Q}(t') = 0$, 
    \item $\comm{\supop{P}(t)}{\frac{d}{dt}} \op{\rho}(t) = 0$.
\end{enumerate}
The last two properties are essential for the derivation of the exact Nakajima-Zwanzig equation for the beyond mean-field dynamics.

We can now derive the formally exact equations of motion for finite-$N$ systems in the form of Nakajima-Zwanzig equations starting from the full Liouville equation,
\begin{align}
    \begin{split}
        \frac{d}{dt} \op{\rho}(t) &= -i \underbrace{\left(\lvl_0 + \sum\limits_{i=1} \lvl_i + \lvl'_i \right)}_{\equiv \lvl} \op{\rho}(t)
        \lvl_0 \op{\rho}.
    \end{split}
\end{align}
We make an expansion in the fluctuations around the mean field dynamics. That is, we split the full Liouvillian into two parts,
\begin{align}
    \lvl_0 + \sum\limits_{i=1} \lvl_i + \lvl'_i &= \underbrace{\left( \left(\lvl_0 + \sum_{i=1} \langle\lvl'_i(t)\rangle_{\text{spins}} \right) + \sum\limits_{i=1} \lvl_i + \langle\lvl'_i(t)\rangle_{0} \right)}_{\text{generator of mean-field dynamics}} + \underbrace{\left( \sum\limits_{i=1} \lvl'_i - \langle\lvl'_i(t)\rangle_{0} - \langle\lvl'_i(t)\rangle_{\text{spins}} \right)}_{\sim\text{ fluctuations}}\nonumber \\
    &=\Bigg( \lvl_0(t)+\sum_{i=1}\lvl_i(t) \Bigg) + \Bigg( \sum_{i=1}\Delta \lvl'_i(t) \Bigg),
\end{align}
where we have defined,
\begin{align}
    \lvl_0(t) &= \lvl_0+\sum_i \langle\lvl'_i(t)\rangle_{\text{spins}}, \\
    \lvl_i(t) &= \lvl_i + \langle\lvl'_i(t)\rangle_{0}, \\
    \Delta \lvl'_i(t) &= \lvl'_i - \langle\lvl'_i(t)\rangle_{0} - \langle\lvl'_i(t)\rangle_{\text{spins}},\\
    \langle\lvl'_i(t)\rangle_{\text{spins}} &=\Tr_i(\lvl_i'\op{\rho}_i(t)), \\
    \langle\lvl'_i(t)\rangle_0 &= \Tr_0(\lvl_i'\op{\rho}_0(t)).
\end{align}

In \cite{WillisPicard1974}, this is termed the ``self-consistent field expansion.'' As for our Dicke model example, this is an expansion valid on both normal and superradiant side. 
Using $\supop{P}(t)$ defined above, we can write down the equation for $\op{\rho}_{\text{rel}}$, assuming we start from an uncorrelated state between photons and all sites,
\begin{align}
    \frac{d\op{\rho}_{\text{rel}}(t)}{dt}=-i\supop{P}(t)\lvl \supop{P}(t)\op{\rho}-\int_0^t dt' \supop{P}(t)\lvl \supop{Q}(t)\supop{G}(t,t')\supop{Q}(t')\lvl \supop{P}(t')\op{\rho}(t'),
    \label{eq:master equation}
\end{align}
where 
\begin{align}
    \supop{G}(t,t') = \tord \exp \left[ -i \int\limits_{t'}^t dt'' \, \supop{Q}(t'') \lvl\right] . 
\end{align}
Using the properties of $\supop{P}(t)$ above, and take either $\Tr_{\overline{0}}$ or $\Tr_{0}\Tr_{\overline{i,j}}$ onto both sides of \newnnref{eq:master equation}, we obtain the following coupled equation of motion,
\begin{align}
    \label{eq:SM-exact-WP-photon} 
    \begin{split}
        \frac{d}{dt} \op{\rho}_0(t) &= -i \lvl_0(t) \op{\rho}_0(t) -\int\limits_0^{t} dt' \, \Tr_{\overline{0}} \left( \sum\limits_{i=1}^{N} \Delta\lvl'_i(t) \supop{G}(t,t') \supop{Q}(t') \sum\limits_{j=1}^N \Delta\lvl'_j(t') (\op{\rho}_0(t') \otimes \op{\rho}_{N,\text{rel}}(t'))\right),
    \end{split} \\
    \begin{split}
       \frac{d}{dt} \op{\rho}_{ij}(t) &= -i (\lvl_i(t)+\lvl_j(t)) \op{\rho}_{ij}(t)- \int\limits_0^t dt' \, \Tr_{\overline{i,j}} \Tr_0 \left( (\Delta\lvl'_i(t) + \Delta\lvl'_j(t)) \supop{G}(t,t') \supop{Q}(t') \sum\limits_{k=1}^N \Delta\lvl'_k(t') (\op{\rho}_0(t') \otimes \op{\rho}_{N,\text{rel}}(t'))\right).
    \end{split} \label{eq:SM-exact-WP-spins}
\end{align}
Since there are already two factors of $\Delta \lvl'$ in the memory term, to lowest order in $\Delta\lvl'$ we can approximate $\supop{G}(t,t')$ by replacing $\lvl$ in the exponential by $\lvl_0(t)+\sum_{i=1}\lvl_i(t)$. By using the property that $[\supop{Q}(t), \lvl_{\alpha}]\supop{Q}(t') = 0$ for any Liouvillian $\lvl_{\alpha}$ acting on a single degree of freedom (either a site or the photon)~\cite{WillisPicard1974}, 
\begin{align}
    \tord \exp \left[ -i \int\limits_{t'}^t dt'' \, \supop{Q}(t'') (\lvl_0(t'')+\sum_{i=1}\lvl_i(t''))\right]\supop{Q}(t')= \tord \exp \left[ -i \int\limits_{t'}^t dt'' \, (\lvl_0(t'')+\sum_{i=1}\lvl_i(t''))\right]\supop{Q}(t'),
\end{align}
we can simplify the calculation and arrive at the equations of motion that describe the evolution of the reduced density matrices $\op{\rho}_0$ and $\op{\rho}_{ij}$.

Now we look at our Dicke model example. We have
\begin{align}
    \begin{split}
        \lvl_0 \op{\rho} &= \comm{\Omega \D{\op{a}}\op{a}}{\op{\rho}} + i\kappa \left( \op{a}\op{\rho}\D{\op{a}} - \frac{\acomm{\D{\op{a}}\op{a}}{\op{\rho}}}{2} \right), \\
        \lvl_i \op{\rho} &= \comm{\omega_z \spin{z}{i}}{\op{\rho}} + (\text{local harmonic bath or Lindbladians}), \\
        \lvl'_i \op{\rho} &= \frac{g}{\sqrt{N}} \comm{(\op{a} + \D{\op{a}})\spin{x}{i}}{\op{\rho}}, \\
        \langle\lvl'_i(t)\rangle_{\text{spins}} \oparg &= \frac{g}{\sqrt{N}} \langle \spin{x}{i}(t) \rangle \comm{\op{a} + \D{\op{a}}}{\oparg}, \\
    \langle\lvl'_i(t)\rangle_0 \oparg &= g \frac{\langle (\op{a} + \D{\op{a}})(t)\rangle}{\sqrt{N}} \comm{\spin{x}{i}}{\oparg}, \\
    \Delta \lvl'_i(t) \oparg&= \frac{g}{\sqrt{N}} \comm{(\op{a} + \D{\op{a}})\spin{x}{i}}{\oparg}-\frac{g}{\sqrt{N}} \langle \spin{x}{i}(t) \rangle \comm{\op{a} + \D{\op{a}}}{\oparg}-g \frac{\langle (\op{a} + \D{\op{a}})(t)\rangle}{\sqrt{N}} \comm{\spin{x}{i}}{\oparg}.
    \end{split}
\end{align}

As a result of expanding the memory kernel to second order in $\Delta \lvl'$, we find that, rather than needing to evolve the full photon density matrix $\op{\rho}_0$, it suffices to keep only moments up to order 2.
For the photon observables, we have for the first moments $q(t) \equiv \langle (\op{a} + \D{\op{a}})(t)\rangle / \sqrt{N}$ and $p(t) \sqrt{N} \equiv \langle -i(\op{a} - \D{\op{a}})(t) \rangle$,
\begin{align}
    \begin{split}
        \frac{d}{dt} q(t) &= \phantom{-}\Omega p(t) - \kappa q(t), \\
        \frac{d}{dt} p(t) &= -\Omega q(t) - \kappa p(t) - 2 g \left( \frac{1}{N} \sum\limits_{k=1}^N \lrangle{\spin{x}{k}(t)} \right).
    \end{split} \label{eq:SM-q_p}
\end{align}
For the second moments, we take the observable fluctuations
\begin{align}
    \begin{split}
        \Delta \lrangle{\D{\op{a}}\op{a}(t)} &= \lrangle{\D{\op{a}}\op{a}(t)} - N \frac{p(t)^2 + q(t)^2}{4}, \\ 
        \Delta \lrangle{(\op{a}+\D{\op{a}})^2(t)} &= \lrangle{(a+\D{\op{a}})^2(t)} - N q(t)^2, \\
        \Delta \lrangle{i(\D{\op{a}}\D{\op{a}}-\op{a}\op{a})(t)} &\equiv \lrangle{i(\D{\op{a}}\D{\op{a}}-\op{a}\op{a})(t)} - N p(t) q(t).
    \end{split}
\end{align}

\begin{align}
\begin{split}
  \frac{d}{dt} \frac{\Delta \lrangle{\D{\op{a}}\op{a} (t)}}{N} &= -2 \kappa \frac{\Delta \lrangle{\D{\op{a}}\op{a} (t)}}{N}, \\
  &\phantom{=} + 2 g^2 \int\limits_0^t dt' \, \Bigg( \frac{1}{N^2} \sum\limits_{i=1}^N \sum\limits_{j=1}^N C_{ij}^{R}(t,t') \Bigg) e^{-\kappa (t-t')} \cos[\Omega(t-t')], \\
  &\phantom{=} + 2 g^2 \int\limits_0^t dt' \, \Bigg( \frac{1}{N} \sum\limits_{i=1}^N \sum\limits_{j=1}^N C_{ij}^{I}(t,t') \Bigg) e^{-\kappa (t-t')} \Bigg( \sin[\Omega(t-t')] \frac{\Delta \lrangle{(\op{a}+\D{\op{a}})^2(t')}}{N}, \\
  &\qquad\qquad\qquad\qquad\qquad\qquad\qquad\qquad\qquad + \cos[\Omega(t-t')] \frac{\Delta \lrangle{i(\op{a}\op{a}-\D{\op{a}}\D{\op{a}})(t')}}{N} \Bigg),
\end{split} \label{eq:SM-adaga_flucs} \\
\begin{split}
  \frac{d}{dt} \frac{\Delta \lrangle{(\op{a}+\D{\op{a}})^2(t)}}{N} &= -2 \Omega \frac{\Delta \lrangle{i(\op{a}\op{a} - \D{\op{a}}\D{\op{a}})(t)}}{N} - 2\kappa \frac{\Delta \lrangle{(\op{a}+\D{\op{a}})^2(t)}}{N} + \frac{2\kappa}{N},
\end{split} \label{eq:SM-square_a+adag_flucs} \\
\begin{split}
  \frac{d}{dt} \frac{\Delta \lrangle{i(\op{a}\op{a}-\D{\op{a}}\D{\op{a}})(t)}}{N} &= 2 \Omega \frac{\Delta \lrangle{(\op{a}+\D{\op{a}})^2(t)}}{N} - 2\kappa \frac{\Delta \lrangle{i(\op{a}\op{a}-\D{\op{a}}\D{\op{a}})(t)}}{N}-\frac{2\Omega}{N}-4\Omega \frac{\Delta \langle \D{\op{a}}\op{a}(t)\rangle}{N} \\
  &\phantom{=} - 4 g^2 \int\limits_0^t dt' \, \Bigg( \frac{1}{N^2} \sum\limits_{i=1}^N \sum\limits_{j=1}^N C_{ij}^{R}(t,t') \Bigg) e^{-\kappa (t-t')}\sin[\Omega(t-t')] \\
  &\phantom{=} - 4 g^2 \int\limits_0^t dt' \, \Bigg( \frac{1}{N} \sum\limits_{i=1}^N \sum\limits_{j=1}^N C_{ij}^{I}(t,t') \Bigg) e^{-\kappa (t-t')} \Bigg( \sin[\Omega(t-t')] \frac{\Delta \lrangle{i(\op{a}\op{a}-\D{\op{a}}\D{\op{a}})(t)}}{N} \\
  &\qquad\qquad\qquad\qquad\qquad\qquad\qquad\qquad\qquad\qquad - \cos[\Omega(t-t')] \frac{\Delta \lrangle{(\op{a}+\D{\op{a}})^2(t')}}{N} \Bigg).
\end{split} \label{eq:SM-aa-adagadag_flucs}
\end{align}
These master equations for the second moments are time-nonlocal due to the memory kernel in \newnnref{eq:SM-exact-WP-photon}.
The time-nonlocality is controlled by correlation functions of the {\it spins}, i.e.,
\begin{align}
\label{eq:SM-spin-spin-corr}
  C_{i,j}^{R}(t,t') + i C_{i,j}^{I}(t,t') &= \Tr_{ij} \Bigg( \left(\atord e^{i \int_{t'}^{t} dt'' \lvl_i(t'') } \spin{x}{i} \right) \left( \Delta \spin{x}{j}(t') \right) \op{\rho}_{i,j}(t') \Bigg),
\end{align}
where $\Delta \spin{x}{i}(t) = \spin{x}{i} - \langle \spin{x}{i}(t) \rangle$.

Finally, the two-site density matrix is propagated via the master equation,
\begin{align}
\label{eq:SM-rhoij}
\begin{split}
  \frac{d}{dt} \op{\rho}_{ij}(t) &= -i \left( \lvl_i + \lvl_j \right) \op{\rho}_{ij} - i \frac{g}{\sqrt{N}} \lrangle{(\op{a}+\D{\op{a}})(t)} \comm{\spin{x}{i} + \spin{x}{j}}{\op{\rho}_{ij}(t)} \\
  &\phantom{=} - \frac{g^2}{N} \int\limits_0^t dt' \, C^R_0(t, t') \comm{\spin{x}{i} + \spin{x}{j}}{\tord e^{-i\int_{t'}^t dt'' \lvl_i(t'') + \lvl_j(t'')} \comm{\Delta \spin{x}{i}(t') + \Delta \spin{x}{j}(t')}{\op{\rho}_{ij}(t')}}  \\
  &\phantom{=} - i \frac{g^2}{N} \int\limits_0^t dt' \, C^I_0(t, t') \Bigg[ \spin{x}{i} + \spin{x}{j}, \tord e^{-i\int_{t'}^t dt'' \lvl_i(t'') + \lvl_j(t'')} \Bigg( \acomm{\Delta \spin{x}{i}(t') + \Delta \spin{x}{j}(t')}{\op{\rho}_{ij}(t')} \\
  &\qquad\qquad + 2(N-2) \Bigg( \Big( \Tr_j \Delta\spin{x}{j}(t') \op{\rho}_{ij}(t') \Big) \otimes \op{\rho}_j(t') + \op{\rho}_i(t') \otimes \Big( \Tr_i \Delta\spin{x}{i}(t') \op{\rho}_{ij}(t') \Big)  \Bigg) \Bigg) \Bigg],
\end{split}
\end{align}
for which the correlation functions entering into the memory kernel are given by photon correlation functions,
\begin{align}
C_{0}^{R}(t,t') + i C_{0}^{I}(t,t') &= \sqrt{N} \Tr_0 \Bigg( \left( \atord e^{i \int_{t'}^{t} dt'' \lvl_0(t'') } (a+\D{\op{a}}) \right) \left( \frac{a + \D{\op{a}}}{\sqrt{N}} - q(t') \right) \op{\rho}_0(t') \Bigg), \\
\begin{split}
    C_{0}^{R}(t,t') &= e^{-\kappa (t-t')} \Bigg( \cos[\Omega (t-t')] \left( \Delta \lrangle{(\op{a}+\D{\op{a}})^2(t')} \right) + \sin[\Omega (t-t')] \left( \Delta \lrangle{i(\D{\op{a}}\D{\op{a}}-\op{a}\op{a})(t')} \right) \Bigg), \\
    C_{0}^{I}(t,t')  &= - e^{-\kappa (t-t')} \sin{[\Omega (t-t')]}.
\end{split} \label{eq:SM-C0}
\end{align}

\subsection{$N$-dependent relaxation time}

In \newnnref{fig:bmf} it was shown that the beyond mean-field equations lead to relaxation times for $\langle \spin{z}{i} \rangle$ proportional to $N/g^2$ for $g < g_c$. 
We sketch here on how this emerges from \newnnref{eq:SM-rhoij}.

Let us start with the mean-field terms, comprising the first line on the right hand side of \newnnref{eq:SM-rhoij}.
These terms should have vanishing contribution, as inherited from the mean-field limit.
The pure dephasing (commuting with $\spin{z}{}$) nature of the local terms $\lvl_i$ and $\lvl_j$ ensure that they will not affect the dynamics of $\langle \spin{z}{i} \rangle$. 
Likewise, at long times ($t\gg 1/\kappa$), the odd-parity quantities $\langle \op{a} + \D{\op{a}} \rangle$ and $\langle \spin{y}{i} \rangle$ will vanish in finite-$N$ systems, which cannot exhibit breaking of the global parity symmetry.
At this point, one already sees that the driver of the long-time dynamics for $\langle \spin{z}{i} \rangle$ must be from the memory terms.
It remains to show that these will indeed have magnitude proportional to $g^2/N$, leading to timescales of $N/g^2$.

To do so, first we make note that the two-site density can be broken up into two parts: the mean-field contribution ($\op{\rho}_i \otimes \op{\rho}_j$) and the irreducible pair correlation. 
While the former leads to observable values $\langle \spin{\mu}{i} \rangle$ and correlations $\langle \spin{\mu}{i}(t) \spin{\mu}{i}(t') \rangle$ that are of order 1, the latter are corrections of order $O(1/N)$.
Not only must this be the case in order for the $N\to\infty$ limit to be exactly given by the factorized ansatz $\op{\rho} = \op{\rho}_0 \otimes \op{\rho}_1 \otimes \cdots \otimes \op{\rho}_N$ at all times, but one sees that this must be self-consistently true in \newnnref{eq:SM-rhoij}, as the time-nonlocal terms are the ones responsible for generating the irreducible correlations.
Thus, the leading order contribution (in $1/N$) to the first time-nonlocal term in \newnnref{eq:SM-rhoij} for $\langle \spin{z}{i} \rangle$ will be
\begin{align}
    - 2i \frac{g^2}{N} \int\limits_0^t dt' \, C^R_0(t, t') \Tr_i \spin{y}{i} \left( \tord e^{-i\int_{t'}^t dt'' \lvl_i(t'')} \comm{\spin{x}{i}(t')}{\op{\rho}_{i}(t')} \right).
\end{align}
Observe that the two-time correlation function will be a quantity that, to leading order, will be independent of $N$.
Finally, in the photon correlation function $C_{0}^R(t,t')$ of \newnnref{eq:SM-C0}, the (unscaled) fluctuations $\langle (\op{a}+\D{\op{a}})^2 \rangle - \langle \op{a}+\D{\op{a}} \rangle^2$ and $\langle i(\D{\op{a}}\D{\op{a}}-\op{a}\op{a}) \rangle - \langle \op{a}+\D{\op{a}} \rangle \langle i(\D{\op{a}}-\op{a}) \rangle$ are of order 1 for $g < g_c$.
Thus this first time-nonlocal term has leading dependence on $N$ as $\propto g^2/N$.

A similar argument follows for the second time-nonlocal term in \newnnref{eq:SM-rhoij}, with $C_{0}^I(t,t')$ clearly being independent of $N$.
Of this, the only part deserving of comment is the term
\begin{align}
    -2i (2i) \frac{g^2}{N} \int\limits_0^t dt' \, C^I_0(t, t') (N-2) \Tr_i \spin{y}{i} \Bigg( \tord e^{-i\int_{t'}^t dt'' \lvl_i(t'')} \Big( \Tr_j \Delta\spin{x}{j}(t') \op{\rho}_{ij}(t') \Big)  \Bigg) .
\end{align}
While there is ostensibly a factor of $N-2$ that would spoil the overall proportionality with $g^2/N$, the separation of $\op{\rho}_{ij}$ into its reducible (mean-field) and irreducible parts show that this is not so.
To wit, the quantity $\Tr_j \Delta\spin{x}{j}(t') \op{\rho}_{ij}(t') \equiv \Tr_j \left( \spin{x}{j} - \langle \spin{x}{j}(t') \rangle \right) \op{\rho}_{ij}(t')$ vanishes when evaluated over the mean-field part of $\op{\rho}_{ij}$.
Thus, such partial traces inherit the $N$ dependence of {\it irreducible} part of $\op{\rho}_{ij}$, which, as we have just argued, must be $O(1/N)$.
This cancels the factor of $N-2$, and in all, one finds that the second time-nonlocal term of \newnnref{eq:SM-rhoij} should be proportional to $g^2/N$.

In the long time limit ($t\gg 1/\kappa$), the evolution of the two-site density matrix can be further approximated as 
\begin{align}
        \frac{d}{dt}\op{\rho}_{ij}(t)&\approx-i(\lvl_i+\lvl_j)\op{\rho}_{ij}(t)-\int\limits_0^{t}K(t,t')\op{\rho}_{ij}(t') \\
        &\approx-i(\lvl_i+\lvl_j)\op{\rho}_{ij}(t)-\int\limits_0^{t}K(t-t')\op{\rho}_{ij}(t') =-i(\lvl_i+\lvl_j)\op{\rho}_{ij}(t)-\int\limits_0^{t}K(t')\op{\rho}_{ij}(t-t') \\
        &\approx-i(\lvl_i+\lvl_j)\op{\rho}_{ij}(t)-\Bigg(\int\limits_0^{\infty}K(t')\Bigg)\op{\rho}_{ij}(t),
\end{align}
where $K(t,t')$ is the memory kernel that's proportional to $g^2/N$. The mean-field term gives contribution to the evolution of $\op{\rho}_{ij}$ of order $g^2/N$ since only the irreducible part of $\op{\rho}_{ij}$ contributes when the system is close to the steady state (terms like $\lvl_i \rho_i(t)$ will vanish at long times on the normal phase side). Going from the first line to the second line, $K(t,t')$ is approximated to be time translationally invariant since when the system is close to the steady state we can ignore time dependence of terms like $\lvl_i(t'')$ and $\Delta \lrangle{(\op{a}+\D{\op{a}})^2(t')}$ in \newnnref{eq:SM-rhoij} and \newnnref{eq:SM-C0}. Finally, we can change the integration upper limit to infinity given that $K(t')$ decays to zero over the time scale $1/\kappa$, which is small compared to $t$. So the evolution becomes approximately Markovian, and for the spin variable $\lrangle{\op{\sigma}_i^z}$ it will decay with a rate proportional to $g^2/N$.

\section{\label{sec:SM-logN-rise-time}Details on logarithmic-in-$N$ dependence of the rise time}
\begin{figure}[h]
    \centering
    \includegraphics[scale=0.75]{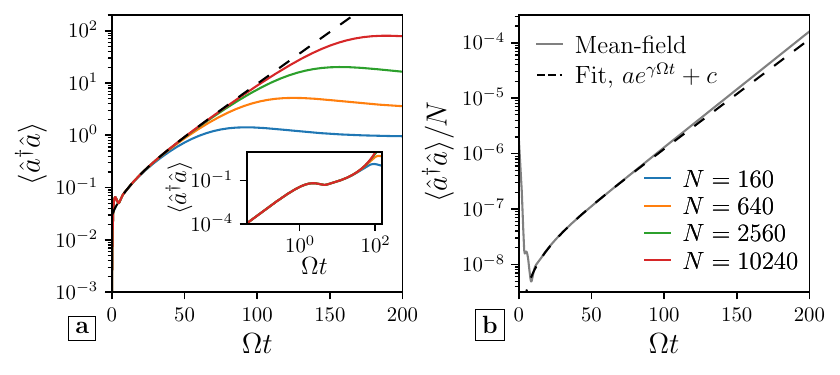}
    \caption{Photon number dynamics for finite $N$ and $N\to\infty$. \textbf{(a)} Growth of the unscaled photon number for $N=160$, $640$, $2560$, and $10240$, starting from an initial state in which the cavity is unoccupied and all spins are fully polarized in the $-\hat{z}$ direction (see \newnnref[c]{fig:bmf}). Dashed line is a fit to $a e^{\gamma \Omega t} + c$. \textbf{(inset)} Same data as (a) but on a log-log plot to show details at early times, $\Omega t \lesssim 1$. It can be seen that all the photon number dynamics coincide across the four values of $N$. \textbf{(b)} Scaled photon number in the mean-field limit. The initial state is the same as in (a), with the small nonzero initial $\langle \op{a} \rangle/\sqrt{N} = 10^{-3}(1+i)$. Like in (a), the dashed line is a fit to $a e^{\gamma \Omega t} + c$ with the same $\gamma \approx 0.045$ in both panels.}
    \label{fig:SM-rise-dynamics}
\end{figure}

In this section, we show data on the early time behavior of the (unscaled) photon number, $\langle \D{\op{a}}\op{a}\rangle$, produced by our beyond-mean-field equations in \newnnref[c]{fig:bmf}.
This observable is shown for $N=160$, $640$, $2560$, and $10240$ in \newnnref[a]{fig:SM-rise-dynamics}.
Additionally, the inset of \newnnref[a]{fig:SM-rise-dynamics} shows the same data on a log-log plot to highlight that the photon number dynamics coincide for all different $N$ for $\Omega t \lesssim 1$.
The dashed line fits the $10 \lesssim \Omega t \leq 50$ data to $a e^{\gamma \Omega t} + c$, from which we obtain $\gamma \approx 0.045$.

In \newnnref[b]{fig:SM-rise-dynamics} we show the mean-field ($N\to\infty)$ scaled photon number, $n(t) \equiv \langle \D{\op{a}}\op{a} \rangle/N$, for the same parameters as in \newnnref[a]{fig:SM-rise-dynamics}.
The initial state is take to be the same as in \newnnref[a]{fig:SM-rise-dynamics}, which has an empty cavity, spins in the $|-\rangle\langle-|$ state, and local baths at zero temperature with respect to $H_B = \sum_j \nu_j\op{b}_j^{\dagger}\op{b}_j$.
Note that this initial condition constitutes an unstable fixed point of the MF dynamics.
To ensure that the dynamics does not get stuck, we perturb the initial photon state such that $\langle a (t=0)\rangle/\sqrt{N} = 10^{-3} + i 10^{-3}$.
The subsequent evolution of the scaled photon number is shown by the solid gray line in \newnnref[b]{fig:SM-rise-dynamics}.
The dashed line shows the fit to $a e^{\gamma \Omega t} + c$ with the same $\gamma$ as in \newnnref[a]{fig:SM-rise-dynamics}, indicating that the behavior shown in \newnnref[a]{fig:SM-rise-dynamics} is inherited from the mean-field dynamics.

The exponential growth shown in these plots should be understood as coming from the largest Lyapunov exponent of the mean-field dynamics evaluated around the initial state.
This exponent is therefore independent of $N$.
Thus, when considering finite $N$ systems, while the dynamics is within the Lyapunov region, we should expect to see this $\exp(\gamma \Omega t)$ growth for $\gamma$ independent of $N$.
Exponentially increasing $\langle \D{\op{a}} \op{a} \rangle$ from $O(N^0)$ to $O(N)$ thus leads to a $O(\log N)$ timescale that we showed in \newnnref[c]{fig:bmf} in the main text.

\section{\label{sec:SM-hierarchical-ordering}Self-consistency of the cluster expansion}

Given the exactness of the mean-field ansatz, $\op{\rho} = \op{\rho}_0 \otimes \op{\rho}_1 \otimes \cdots \otimes \op{\rho}_N$, for the model under consideration~\cite{Mori2013}, it is necessary that our approach to beyond-mean-field dynamics (\newnnref{sec:SM-bmf}) be able to recover the MF limit as $N\to\infty$.
As our approach is based on projecting the full state on to a cluster expansion including up to pairwise site correlations,
\begin{align*}
    \op{\rho}_{\text{rel}} &= \op{\rho}_0 \otimes \Big[ \underbrace{\bigotimes\limits_{i=1}^N \op{\rho}_i}_{\text{``MF''}} + \underbrace{\sum\limits_{i>j} (\op{\rho}_{ij} - \op{\rho}_i \otimes \op{\rho}_j)\otimes \bigotimes\limits_{k\neq i, j}^N \op{\rho}_k}_{\text{Irred.\ 2-site corrs.}} \Big],
\end{align*}
in order to properly recover the MF limit, we must have that the irreducible two-site correlations vanish as $N\to\infty$.

To characterize this, we show how the connected part of the two site correlation $\langle \spin{\mu}{i} \spin{\nu}{j} \rangle - \langle \spin{\mu}{i} \rangle \langle \spin{\nu}{j} \rangle$ at a fixed time changes with increasing $N$.
\begin{figure}[h]
    \centering
    \includegraphics[scale=0.75]{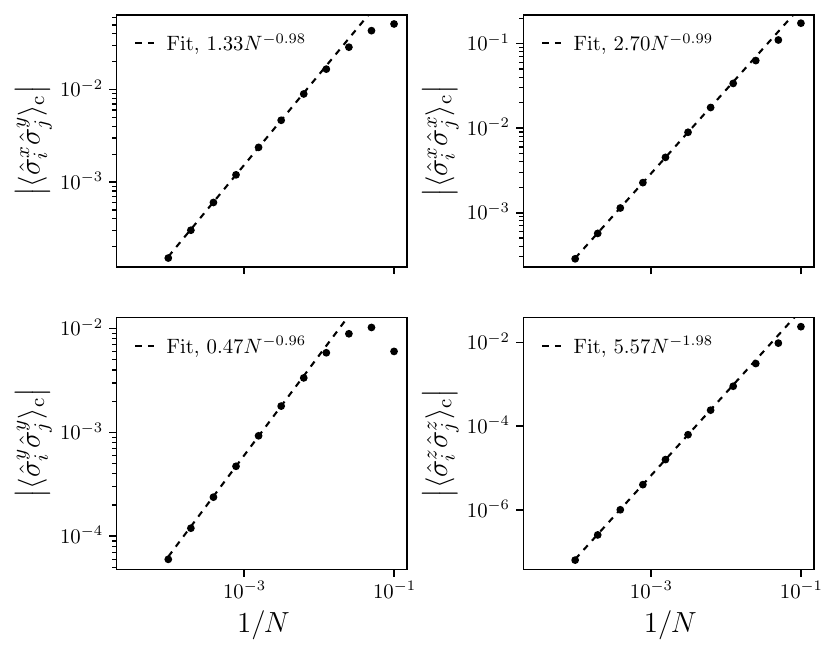}
    \caption{Decay with $N$ of the magnitude of the connected two-site spin correlators, $\langle \spin{\mu}{i} \spin{\nu}{j} \rangle_c \equiv \langle \spin{\mu}{i} \spin{\nu}{j} \rangle - \langle \spin{\mu}{i} \rangle\langle \spin{\nu}{j} \rangle$, evaluated at $\Omega t = 20.0$, with all spins initially in the state $\op{\rho}_i(0) = |\downarrow\rangle\langle\downarrow|$. Since the dynamics obeys parity symmetry, we show only the four nonzero correlators. Shown in the dashed lines are power law fits to the large $N$ behavior of the connected correlations. The parameters here are the same as the ones for \newnnref[c]{fig:bmf} in the main text, namely that $g = 0.26 > g_c$.}
    \label{fig:SM-connected-corrs-SR}
\end{figure}

\begin{figure}[h]
    \centering
    \includegraphics[scale=0.75]{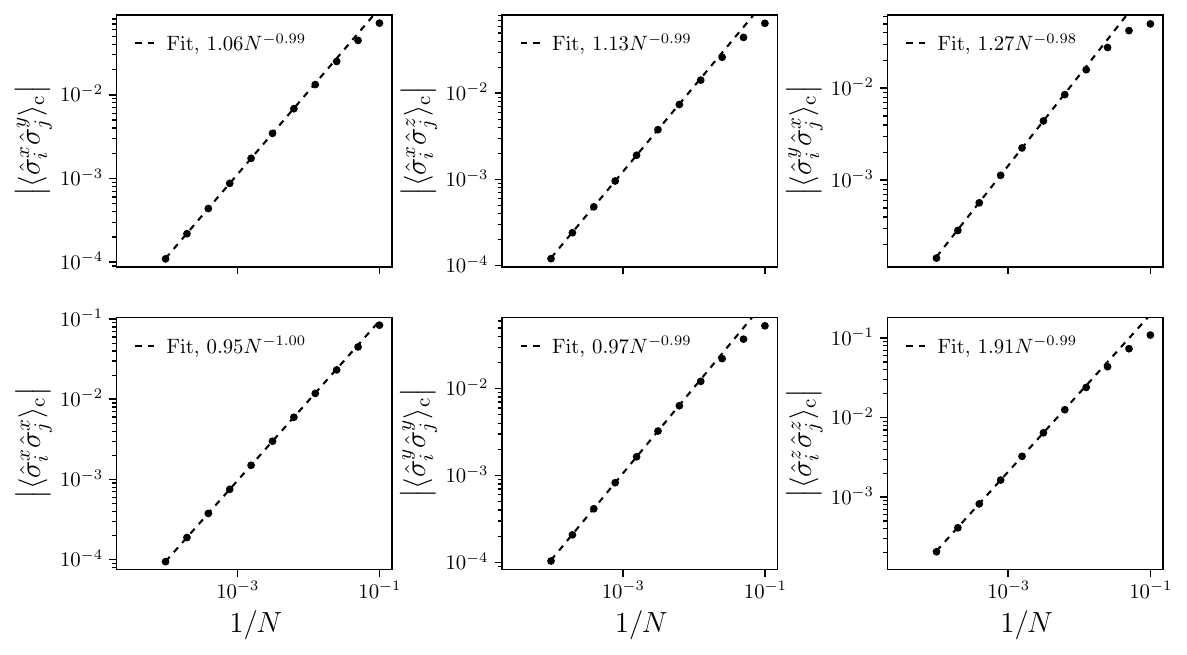}
    \caption{Same as in \newnnref{fig:SM-connected-corrs-SR}, but with the spins initially tilted at an angle of $\pi/8$ away from the $+\hat{z}$ direction, towards the $+\hat{x}$ axis. Since the dynamics is not restricted to the parity symmetric subspace, we show all six possible connected correlations.}
    \label{fig:SM-connected-corrs-SR-tilted}
\end{figure}

From \newnnref{fig:SM-connected-corrs-SR} and \newnnref{fig:SM-connected-corrs-SR-tilted}, we see that for a fixed time, as $N$ is increased, all connected correlations vanish at least as fast as $1/N$.
Hence if one were to take $N\to\infty$ before $t\to\infty$, one will find that the mean-field result holds for all times.
This supports the idea that our beyond-mean-field equations is self-consistent, in that it can recover the known exact mean-field limit.
We show explicitly in \newnnref{sec:SM-naive-projector-bad} an example for which one cannot recover the mean-field limit, as a result of a bad choice of projector and an insufficient expansion of the memory kernel.

\section{\label{sec:SM-naive-projector-bad}Inadequacy of the Born approximation using the na\"ive projection operator}

We choose the projection operator to be
\begin{align}
\supop{P}(t)\oparg=-N \Tr(\oparg)\overset{N}{\underset{n=0}{\otimes}}\op{\rho}_n(t)+\sum_{m=0}^N \Tr_{\overline{m}}(\oparg)\otimes \Big(\overset{N}{\underset{\substack{n\neq m\\n=0}}{\bigotimes}} \op{\rho}_{n}(t)\Big),
\end{align}
such that 
\begin{align}
    \op{\rho}_{\text{rel}}(t)=\supop{P}(t)\op{\rho}(t)=\overset{N}{\underset{n=0}{\otimes}}\op{\rho}_n(t),
\end{align}
and that $\supop{P}(t)$ obeys the same five properties as in \newnnref{sec:SM-bmf}.
Following the same procedure as in Sec~\ref{sec:SM-bmf}, which makes use of a Born approximation by expanding the time-nonlocal memory kernel to second order in the coupling, we obtain the evolution equations of $\op{\rho}_0(t)$ and $\op{\rho}_k(t)$,
\begin{align}
    \begin{split}
    \frac{d}{dt} \op{\rho}_0(t) &= -i \lvl_0 \op{\rho}_0(t) -i g \sqrt{N} \left( \frac{\sum_k\langle \spin{x}{k}(t)\rangle}{N} \right) \comm{\op{a} + \D{\op{a}}}{\op{\rho}_0(t)} \\
&-g^2\int_0^t dt'\Bigg(\frac{1}{N}\sum_{k=1}^N C_{kk}^R(t,t')\Bigg)
\comm{\op{a}+\D{\op{a}}}{ \tord e^{-i\int_{t'}^t dt'' \lvl_0(t'')} \comm{\op{a}+\D{\op{a}}-\langle(\op{a}+\D{\op{a}})(t')\rangle}{\op{\rho}_0(t')}} \\
&-g^2\int_0^t dt'\Bigg(\frac{i}{N}\sum_{k=1}^N C_{kk}^I(t,t')\Bigg)
\comm{\op{a}+\D{\op{a}}}{ \tord e^{-i\int_{t'}^t dt'' \lvl_0(t'')} \acomm{\op{a}+\D{\op{a}}-\langle(\op{a}+\D{\op{a}})(t')\rangle}{\op{\rho}_0(t')}},
    \end{split} \\
    \begin{split}
\frac{d}{dt}\op{\rho_k}(t)
&=-i \lvl_k  \op{\rho}_{k}(t) - i g \frac{\langle (\op{a} + \D{\op{a}}) (t) \rangle}{\sqrt{N}} \comm{\spin{x}{k}}{\op{\rho}_{k}(t)} \\
&-\frac{g^2}{N}\int_0^t dt'C^R_0(t, t') \comm{\spin{x}{k}}{\tord e^{-i\int_{t'}^t dt'' \lvl_k(t'')} \comm{\Delta \spin{x}{k}(t')}{\op{\rho}_{k}(t')}} \\
&-\frac{ig^2}{N}\int_0^t dt'C^I_0(t, t') \comm{\spin{x}{k}}{\tord e^{-i\int_{t'}^t dt'' \lvl_k(t'')} \acomm{\Delta \spin{x}{k}(t')}{\op{\rho}_{k}(t')}}.
\end{split}
\end{align}
Here, the correlation functions (e.g.\ $C^R_0$, $C^I_{kk}$, etc.) are the same as those defined in \newnnref{eq:SM-spin-spin-corr}.
Notice that using this naive projector, the memory kernel of the spin density matrix will only depend on spin-spin correlation functions that involve spins at the same site, in contrast to the $C_{ij}^R(t,t')$ in the previous section which could involve spins at different sites.

We will again see that it suffices to keep the first and second moments of photon density matrix. For simplicity we look at the case without the local bath, for which we know there exists a superradiant phase transition, and we will obtain coupled equations of five photon variables as well as three spin variables $\langle \sigma_k^x\rangle$, $\langle \sigma_k^y\rangle$ and $\langle \sigma_k^z\rangle$. If we focus on the steady state and ignore all the symmetry breaking terms for simplicity, 
\begin{align}
\begin{split}
  0=\frac{d}{dt} \frac{ \lrangle{\D{\op{a}}\op{a} (t\rightarrow \infty)}}{N} &= -2 \kappa \frac{\lrangle{\D{\op{a}}\op{a} (t\rightarrow \infty)}}{N} \\
  &\phantom{=} + 2 g^2 \int\limits_0^\infty dt' \, \Bigg(\frac{1}{N^2}\sum_{k=1}^N C_{kk}^R(t')\Bigg) e^{-\kappa t'} \cos[\Omega t'] \\
  &\phantom{=} + 2 g^2 \int\limits_0^\infty dt' \, \Bigg(\frac{1}{N}\sum_{k=1}^N C_{kk}^I(t')\Bigg) e^{-\kappa t'} \Bigg( \sin[\Omega t'] \frac{ \lrangle{(\op{a}+\D{\op{a}})^2(t\rightarrow \infty)}}{N} \\
  &\qquad\qquad\qquad\qquad\qquad\qquad\qquad\qquad\qquad + \cos[\Omega t'] \frac{ \lrangle{i(\op{a}\op{a}-\D{\op{a}}\D{\op{a}})(t\rightarrow \infty)}}{N} \Bigg),
\end{split} \label{eq:SM-adaga} \\
\begin{split}
  0=\frac{d}{dt} \frac{ \lrangle{(\op{a}+\D{\op{a}})^2(t\rightarrow \infty)}}{N} &= -2 \Omega \frac{ \lrangle{i(\op{a}\op{a} - \D{\op{a}}\D{\op{a}})(t\rightarrow\infty)}}{N} - 2\kappa \frac{ \lrangle{(\op{a}+\D{\op{a}})^2(t\rightarrow \infty)}}{N} + \frac{2\kappa}{N},
\end{split} \label{eq:SM-square_a+adag} \\
\begin{split}
  0=\frac{d}{dt} \frac{ \lrangle{i(\op{a}\op{a}-\D{\op{a}}\D{\op{a}})(t\rightarrow \infty )}}{N} &= 2 \Omega \frac{\lrangle{(\op{a}+\D{\op{a}})^2(t\rightarrow \infty)}}{N} - 2\kappa \frac{ \lrangle{i(\op{a}\op{a}-\D{\op{a}}\D{\op{a}})(t\rightarrow \infty)}}{N}-\frac{2\Omega}{N}-4\Omega \frac{\langle \D{\op{a}}\op{a}(t\rightarrow \infty)\rangle}{N} \\
  &\phantom{=} - 4 g^2 \int\limits_0^\infty dt' \, \Bigg(\frac{1}{N^2}\sum_{k=1}^N C_{kk}^R(t')\Bigg) e^{-\kappa t'}\sin[\Omega t'] \\
  &\phantom{=} - 4 g^2 \int\limits_0^\infty dt' \, \Bigg(\frac{1}{N}\sum_{k=1}^N C_{kk}^I(t')\Bigg) e^{-\kappa t'} \Bigg( \sin[\Omega t'] \frac{ \lrangle{i(\op{a}\op{a}-\D{\op{a}}\D{\op{a}})(t\rightarrow \infty)}}{N} \\
  &\qquad\qquad\qquad\qquad\qquad\qquad\qquad\qquad\qquad\qquad - \cos[\Omega t'] \frac{ \lrangle{(\op{a}+\D{\op{a}})^2(t\rightarrow \infty)}}{N} \Bigg),
\end{split} \label{eq:SM-aa-adagadag} \\
\begin{split}
    0=\frac{d}{dt}\lrangle{\sigma_k^z(t\rightarrow \infty)}&=-\frac{g^2}{N}\int_0^{\infty} 4 C_0^R(t')\cos(2\omega_zt')\lrangle{\sigma_k^z(t\rightarrow \infty)} +\frac{g^2}{N}\int_0^{\infty}dt'4C_0^I(t')\sin(2\omega_zt'),
\end{split}
\end{align}
where following the notation in Section~\ref{sec:SM-bmf}, the correlation functions are defined as
\begin{align}
    \begin{split}
        C_{kk}^R(t') &= \cos(2\omega_zt'), \\
        C_{kk}^I(t') &= \sin(2\omega_zt')\langle \sigma^z_k(t\rightarrow \infty)\rangle,
    \end{split}   \\
    \begin{split}
         C_{0}^{R}(t') &= e^{-\kappa t'} \Bigg( \cos(\Omega t') \left(  \lrangle{(\op{a}+\D{\op{a}})^2(t\rightarrow \infty)} \right) + \sin(\Omega t') \left(  \lrangle{i(\D{\op{a}}\D{\op{a}}-\op{a}\op{a})(t\rightarrow \infty)} \right) \Bigg), \\
    C_{0}^{I}(t')  &= - e^{-\kappa t'} \sin{(\Omega t')}. 
    \end{split}
\end{align}

Under permutation invariance, we can see that the steady state equations are all proportional to $1/N$, so that all the $N$ dependence can be trivially removed, leaving us with $N$ independent unscaled steady state photon number. Therefore the scaled steady state photon number will drop to zero as $N$ goes to infinity, indicating the absence of the superradiant phase transition. Even if we keep the symmetry breaking terms in the above equations, the situation does not change and we still obtain the same $N$ independent solution. Thus, one would either need to expand the memory kernel to higher order when using the na\"ive projector, or to take a different projector. A similar observation was made by Degenfeld-Schonburg in his thesis~\cite{DegenfeldSchonburgThesis}.

\section{\label{sec:SM-benchmark}Benchmark with respect to the Dicke model without the local baths}
\begin{figure}[h]
        \centering
        \begin{tabular}{cc}
            \includegraphics[width=0.45\linewidth]{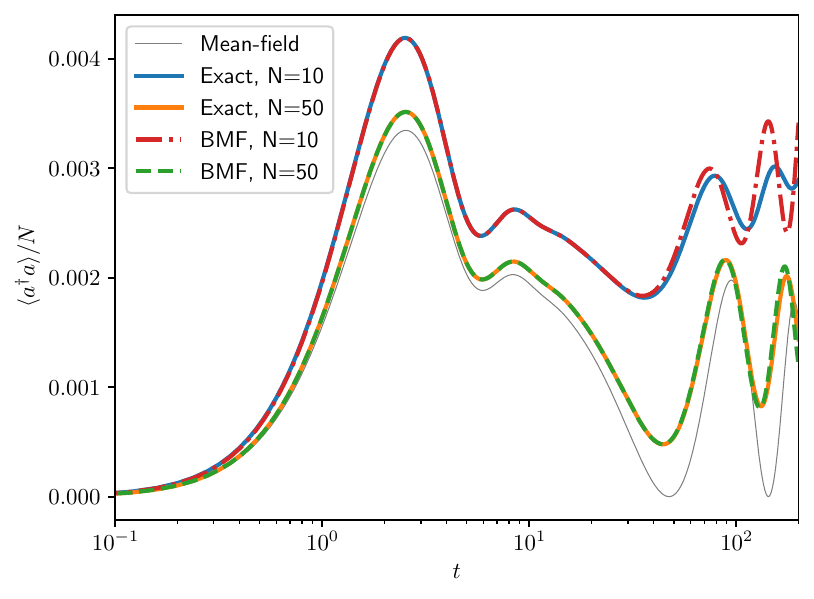} & \includegraphics[width=0.45\linewidth]{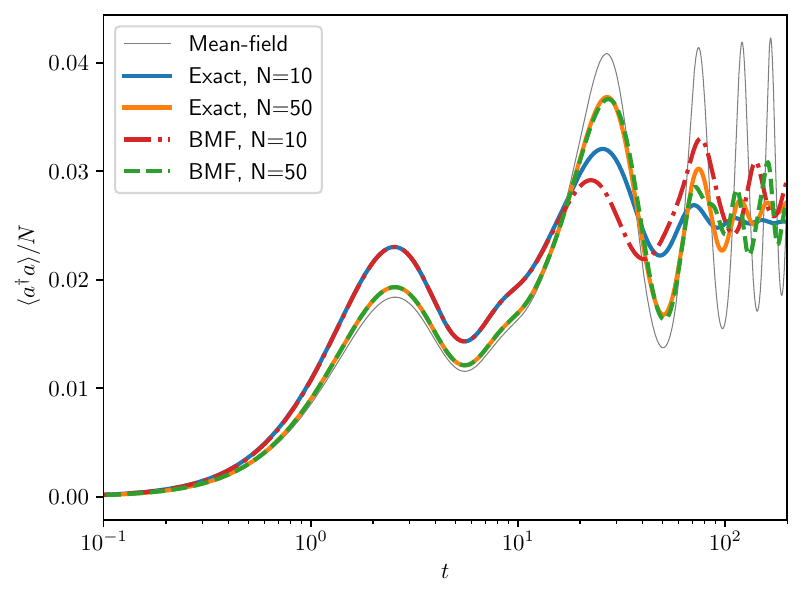} \\
            $g = 0.1$ & $g = 0.234$
        \end{tabular}
        \caption{Benchmark comparison of the exact dynamics (solid lines) versus the result of our beyond mean field (BMF) method (dashed and dash dotted lines), for $N = 10, 50$ spins without any local baths. Here we give the dynamics in the Dicke model for two values of the light-matter coupling. For reference, the exact dynamics in the mean field ($N\to\infty$) limit are provided in the thin black lines. As $N$ increases, not only do the BMF dynamics agree with the exact dynamics for longer times, but it also tends towards the limiting mean field solution. All other parameters are the same
        as in the main text.}
        \label{fig:SM-comparison-tilted}
    \end{figure}

In Fig.~\ref{fig:SM-comparison-tilted} we present results comparing our beyond mean field method (BMF) against numerical exact diagonalization and the exact mean field limit using an initial condition where the cavity is in the vacuum state and spins are initially polarized in the $\hat{x}-\hat{z}$ plane at an angle of $\theta = 0.825\pi$ from the $+\hat{z}$ direction. We can clearly see that as $N$ increases, our beyond mean field method (`BMF') is consistent with the exact result to much longer time, and both the approximate and exact dynamics approach the mean field limit. This behavior is expected because our beyond mean field method is a $1/N$ expansion around the mean field dynamics, and in the thermodynamic limit it will reduce to the mean field result.

\section{\label{sec:SM-benchmark-singleboson}Benchmark with respect to the local baths modeled by a single bosonic mode with Lindbladian dissipation}

\begin{figure}[h]
        \centering
        \begin{tabular}{cc}
            \includegraphics[width=0.45\linewidth]{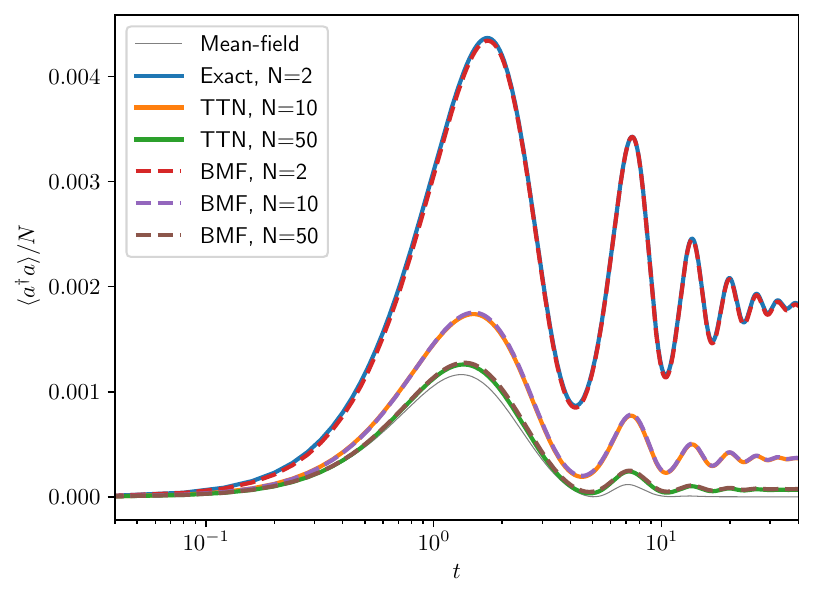} & \includegraphics[width=0.45\linewidth]{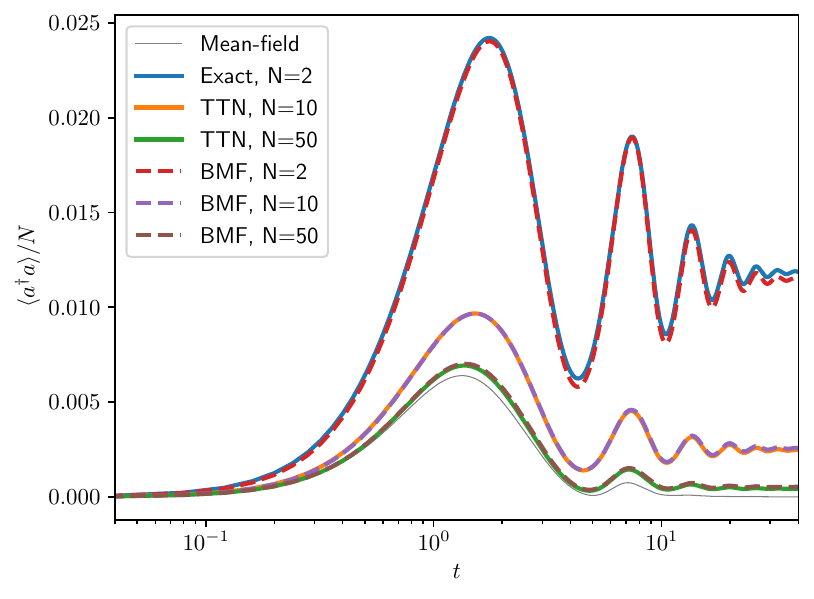} \\
            $g = 0.1$ & $g = 0.234$
        \end{tabular}
        \caption{Benchmark with a local bath modeled by a single bosonic mode undergoing Lindbladian dissipation.}
        \label{fig:1boson-comparison-tilted}
        \end{figure}

To check the accuracy of our formalism when the two-level systems are themselves coupled to any additional degree of freedom acting as a bath, we examine the same Dicke model where the Hamiltonian for the $n$-th site is
        \begin{align}
            H_{n} &= \omega_z \spin{z}{n} + \sqrt{\alpha} \spin{x}{n} (b + b^{\dagger}) + \omega_b b^{\dagger} b.
        \end{align}
Each spin couples to its own bosonic mode of frequency $\omega_b$.
        The bosonic mode is subjected to Lindbladian dissipation, $\mathcal{D}[b] \oparg = \gamma (b \oparg b^{\dagger} - \acomm{b^\dagger b}{\oparg}/2)$.
        This is effectively the same as coupling each TLS to a continuous bosonic bath with a Lorentzian spectral density \cite{Tamascelli2018}.
        In Fig.~\ref{fig:1boson-comparison-tilted}, we present a benchmark of our method (``BMF'') against an exact calculation for $N=2$ using QuTiP \cite{johansson2012qutip} and additional calculations for $N=10, 50$ using a recently released tree-tensor network (TTN) package \text{pyTTN} \cite{Lindoy2025}.         
        Here, we take the same cavity and spin parameters as in our manuscript. 
        The bosonic mode and its coupling are specified by $\alpha = 0.3, \omega_b = 1, \gamma = 0.25$, and the maximum occupation of the mode is truncated to 5.
        The mode is initially unoccupied.
        Across all three methods, the bath is represented in the same way so that we are only testing the validity of the BMF formalism. Based on Fig.~\ref{fig:1boson-comparison-tilted}, we can see that our BMF results are in good agreement with the exact and TTN results. 
    
\section{\label{sec:SM-benchmark-ohmicbath}Benchmark with respect to the local Ohmic baths}
\begin{figure}[h]
        \centering
        \begin{tabular}{cc}
            \includegraphics[width=0.45\linewidth]{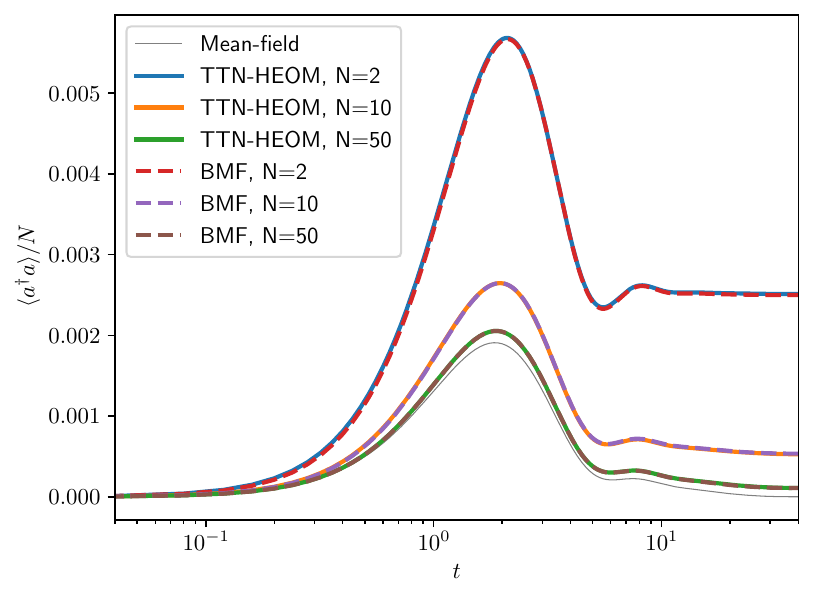} & \includegraphics[width=0.45\linewidth]{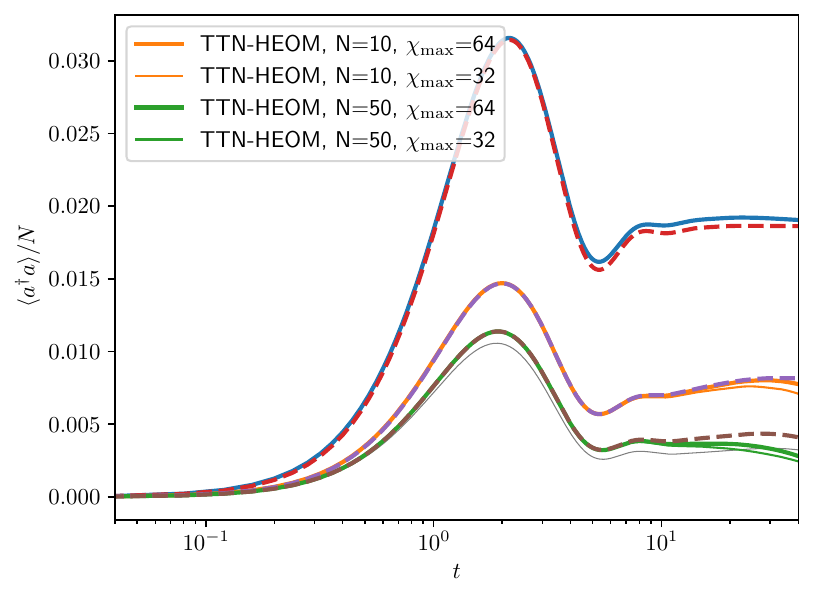} \\
            $g = 0.1$ & $g = 0.234$
        \end{tabular}
        \caption{Benchmark with the local bath used in the manuscript, an exponentially cut-off Ohmic spectral density. Here we compare against our BMF dynamics using the iTEBD-TEMPO representation of the bath against a direct tree-tensor network calculation with a HEOM representation of the bath.}
        \label{fig:heom-comparison-tilted}
        \end{figure}

Having demonstrated that our formalism is valid with the addition of a physical local bath, we now turn to check the accuracy of our representation of the bath.
        In all of our calculations in the manuscript, unless otherwise indicated, we have taken the local bosonic bath to have an exponentially cut-off Ohmic spectral density.
        The bath is represented using the time-translationally invariant formulation of the Time-Evolving Matrix Product Operator (iTEBD-TEMPO) method \cite{Link2024}.
        In Fig.~\ref{fig:heom-comparison-tilted} we compare our BMF dynamics with the Ohmic local bath to TTN calculations with a HEOM representation of the local bath using \text{pyTTN}, again for $N=2, 10, 50$.
        The HEOM calculations are performed with a bath decomposition into $K=10$ exponential modes using the ESPRIT algorithm to fit the bath-bath correlation function. 
        The ADO hierarchy is truncated such that the occupation of the bosonic modes is at most $L=20$.
        Finally, the TTN 
        has a maximum bond dimension of $\chi_{\text{max}} = 64$ unless otherwise stated.

        While the methods compare well in the weaker coupling case, we see that at larger coupling there are larger disagreements at longer time for ``large'' $N$, most evident for $N=50$. We believe this is because the TTN-HEOM calculations are not converged with respect to the bond dimension $\chi_{\text{max}}$.
        We show by comparison to calculations with smaller bond dimension $\chi_{\text{max}} = 32$ (thin lines) that there is still a drift in the TTN calculations, indicating poor convergence for times $t \gtrsim 10$.
        However, the trend appears to be that the TTN-HEOM results are converging towards the BMF results (see especially the result for $N=10$ for $g = 0.234$).

\section{\label{sec:SM-benchmark-finiteT}Finite temperature local baths}
\begin{figure}[h]
            \hspace*{-0.6in}
            \begin{tabular}{ccc}
                 \includegraphics[width=0.38\textwidth]{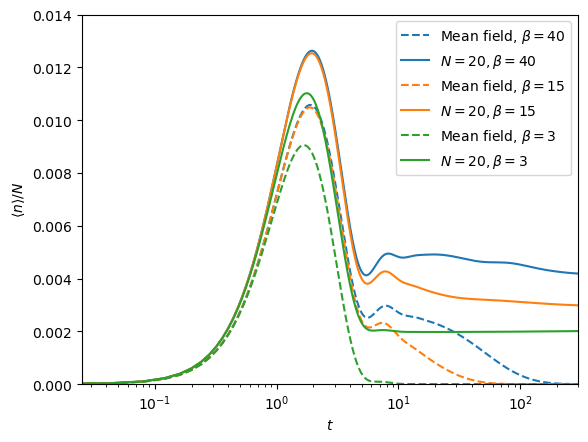} & \includegraphics[width=0.38\textwidth]{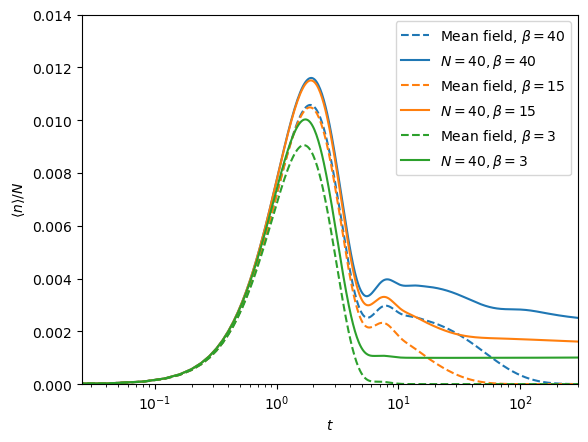} & \includegraphics[width=0.38\textwidth]{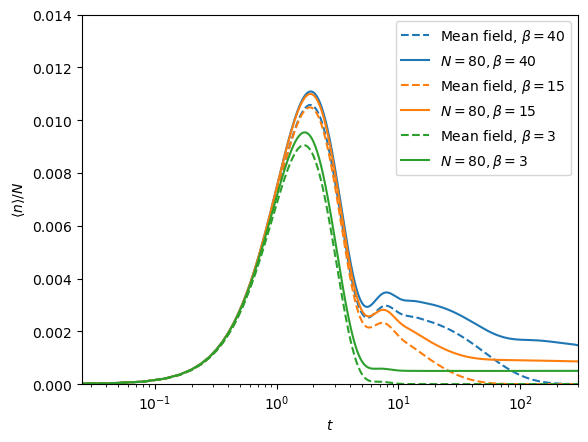} \\
                 (a) $N=20$ vs. MF & (b) $N=40$  vs. MF & (c) $N=80$ vs. MF
            \end{tabular}
            \caption{$g \approx 0.234$ with the same initial condition across all cases}
            \label{fig:finite-temp}
        \end{figure}

Since our formalism is based on density matrices, it's applicable to finite temperature local baths. In Fig.~\ref{fig:finite-temp} we show the scaled photon number for $N = 20, 40, 80$ across three panels, with each comparing the finite-temperature MF and beyond MF dynamics on a semi-log scale.

 We show results for three temperatures:
        \begin{itemize}
            \item $k T = 1/40 < \Delta$: Thermal energy is low in comparison to the Zeeman splitting $\Delta \equiv 2\omega_z$ for the two-level system;
            \item $\Delta < k T = 1/15 < \lambda$: Thermal energy is between the Zeeman splitting $\Delta$ and the reorganization energy $\lambda \equiv \alpha \omega_c/2$ for the local dephasing bath;
            \item $\Delta, \lambda, g < k T = 1/3 < \Omega, \kappa$: Thermal energy is greater than the Zeeman splitting, local bath reorganization energy and the strength $g$ of collective cavity-spin coupling, while being less than the cavity frequency or the photon loss rate.
        \end{itemize}  
        
While higher temperature does have an effect on the photon number, it is evident that they do not wash out the deviation from MF, which is due to the finite $N$ effect.

\end{document}